\documentclass[11pt]{article}

\usepackage{graphicx}
\usepackage{amsmath}
\usepackage{dcolumn}
\usepackage{bm}
\usepackage{slashed}
\usepackage[bookmarks,bookmarksnumbered,colorlinks=true,anchorcolor=blue,
linkcolor=blue,urlcolor=blue,citecolor=blue,breaklinks=true]{hyperref}
\usepackage[utf8]{inputenc}
\usepackage{authblk}
\usepackage{cite}
\usepackage{titlesec}
\usepackage{caption}
\usepackage{subcaption}

\usepackage{color}
\usepackage{ulem}

\newcommand{\be}{\begin{equation}}
\newcommand{\ee}{\end{equation}}
\newcommand{\ba}{\begin{eqnarray}}
\newcommand{\ea}{\end{eqnarray}}
\def\bea{\begin{eqnarray}}
\def\eea{\end{eqnarray}}

\newcommand{\gsim}{\mathrel{\hbox{\rlap{\lower.55ex \hbox {$\sim$}}
                   \kern-.3em \raise.4ex \hbox{$>$}}}}
\newcommand{\lsim}{\mathrel{\hbox{\rlap{\lower.55ex \hbox {$\sim$}}
                   \kern-.3em \raise.4ex \hbox{$<$}}}}

\def\roughly#1{\mathrel{\raise.3ex\hbox{$#1$\kern-.75em%
\lower1ex\hbox{$\sim$}}}}
\def\lsim{\roughly<}
\def\gsim{\roughly>}

\def\({\left(}
\def\){\right)}
\def\[{\left[}
\def\]{\right]}
\def\<{\langle}
\def\>{\rangle}

\usepackage{xcolor}

\begin{document}
\title{\bf Holographic coarse-grained states and the necessity of perfect entanglement }

\author[]{Yi-Yu Lin$^{1,2}$ \thanks{yiyu@bimsa.cn}}
\author[]{Jun Zhang$^{3}$ \thanks{jzhang163@crimson.ua.edu}}

\affil[]{${}^1$Beijing Institute of Mathematical Sciences and Applications (BIMSA),
	Beijing, 101408, China}
 \affil{${}^2$Yau Mathematical Sciences Center (YMSC), Tsinghua University,
	Beijing, 100084, China}
  \affil{${}^3$Department of Physics and Astronomy, University of Alabama, 514 University Boulevard, Tuscaloosa, AL 35487, USA}


\maketitle

\begin{abstract}

In the framework of the holographic principle, focusing on a central concept, conditional mutual information, we construct a class of coarse-grained states, which are intuitively connected to a family of thread configurations. These coarse-grained states characterize the entanglement structure of holographic systems at a coarse-grained level. Importantly, these coarse-grained states can be used to further reveal nontrivial requirements for the holographic entanglement structure. Specifically, we employ these coarse-grained states to probe the entanglement entropies of disconnected regions and the entanglement wedge cross-section dual to the inherent correlation in a bipartite mixed state. The investigations demonstrate the necessity of perfect tensor state entanglement. Moreover, in a certain sense, our work establishes the equivalence between the holographic entanglement of purification and the holographic balanced partial entropy. We also construct a thread configuration with the Multi-Scale Entanglement Renormalization Ansatz (MERA) structure, reexamining the connection between the MERA structure and kinematic space.

\end{abstract}
\tableofcontents

\newpage

\section{Introduction}

In the framework of holographic duality~\cite{Maldacena:1997re,Gubser:1998bc,Witten:1998qj}, the Ryu-Takayanagi (RT) formula~\cite{Ryu:2006bv,Ryu:2006ef,Hubeny:2007xt} for holographic entanglement entropy suggests a profound connection between gravity and quantum entanglement. The formula states that the entanglement entropy $S(A)$ of a subsystem  $A$ in the holographic quantum system can be precisely calculated by the area of a minimal extremal surface $\gamma_A$ in the dual higher-dimensional spacetime:
\be\label{rt}
S_A = \frac{{\text{{Area}}(\gamma_A)}}{{4G_N}},\ee
where $\gamma_A$ is homologous to $A$ and completely separating $A$ from $B$. 

Following this clue, tensor networks have proven useful for studying the entanglement structure in holographic gravity~\cite{Vidal:2007hda,Vidal:2008zz,Vidal:2015,Swingle:2009bg,Swingle:2012wq,Pastawski:2015qua,Hayden:2016cfa,Chen:2021ipv,Chen:2021qah,Chen:2021rsy,Bao:2018pvs,Bao:2019fpq,Haegeman:2011uy,Qi:2013caa,Miyaji:2015fia,Miyaji:2015yva,Chen:2022wvy,Almheiri:2014lwa}.
~\footnote{For more research on tensor networks in the holographic context, see e.g.~\cite{Czech:2015xna,Evenbly:2017hyg,Steinberg:2023wll,Steinberg:2020bef,Czech:2016nxc,Singh:2017tet,Colafranceschi:2022ual,Cheng:2023kxh,Zeng:2023dzh, Hung:2019zsk, Milsted:2018vop,Milsted:2018yur, Milsted:2018san, SinaiKunkolienkar:2016lgg,Bao:2017qmt,Beny:2011vh,Czech:2015kbp, Ling:2019akz,Ling:2018ajv,Ling:2018vza,Bhattacharyya:2017aly,Bhattacharyya:2016hbx, Gan:2017nyt,Bao:2015uaa,Yu:2020zwk,Sun:2019ycv,Belin:2023efa,Lin:2020ufd}.} This tool was initially employed in the intersection of condensed matter physics and quantum information theory to characterize a series of states (especially ground states) of quantum many-body systems. Essentially, they describe a class of states that can be represented as a contraction of many small tensors. Each small tensor is graphically represented as a subdiagram with legs extending from a vertex, while tensor contractions are represented by connections between legs of different subdiagrams. Finally, these subdiagrams are connected into a network pattern, known as a tensor network. Our research is interested in characterizing the entanglement structure of holographic gravity from a complementary perspective—the ``thread" perspective~\cite{Freedman:2016zud,Cui:2018dyq,Headrick:2017ucz,Headrick:2022nbe,Headrick:2020gyq,Lin:2020yzf,Lin:2021hqs,Lin:2022aqf,Lin:2022flo,Lin:2022agc,Lin:2023orb}. Similarly, the thread perspective is also geometrically intuitive. The difference is that, in the tensor network picture, the objects of interest are tensors represented by local subdiagrams, while in the thread perspective, the objects of interest are threads with global features. Broadly speaking, we can define a thread configuration as a collection of threads, wherein the endpoints of the threads are anchored in the boundary quantum system. This perspective is inspired by the so-called bit threads~\cite{Freedman:2016zud,Cui:2018dyq,Headrick:2017ucz,Headrick:2022nbe} ~\footnote{For the recent developments of bit threads see e.g.~\cite{Kudler-Flam:2019oru,Rolph:2021nan,Harper:2022sky,Harper:2021uuq,Lin:2022aqf,Lin:2022flo,Lin:2022agc,Lin:2023orb,Lin:2023rbd,Lin:2021hqs,Lin:2020yzf,Headrick:2020gyq,Agon:2021tia,Rolph:2021hgz,Chen:2018ywy,Hubeny:2018bri,Agon:2018lwq,Du:2019emy,Bao:2019wcf,Harper:2019lff,Agon:2019qgh,Du:2019vwh,Agon:2020mvu,Bao:2020uku,Pedraza:2021fgp,Pedraza:2021mkh,Harper:2018sdd,Shaghoulian:2022fop,Susskind:2021esx,Bakhmatov:2017ihw}.}, which arise from the rephrasing of the RT formula for holographic entanglement entropy. 

In a series of previous works~\cite{Lin:2021hqs,Lin:2022aqf,Lin:2022flo,Lin:2022agc,Lin:2023orb}, we constructed a class of thread configurations closely related to many concepts in the research of holographic duality, such as bit threads~\cite{Freedman:2016zud,Cui:2018dyq,Headrick:2017ucz,Headrick:2022nbe}, kinematic space~\cite{Czech:2015kbp,Czech:2015qta}, holographic entropy cone~\cite{Bao:2015bfa,Hubeny:2018ijt,Hubeny:2018trv,HernandezCuenca:2019wgh}, and holographic partial entanglement entropy~\cite{Vidal:2014aal,Wen:2019iyq,Wen:2018whg,Wen:2020ech,Han:2019scu,Kudler-Flam:2019oru,Lin:2021hqs}. These connections are somewhat natural and easily obtained. Especially, the conditional mutual information (CMI), which characterizes the density of entanglement entropy in some sense, plays a central role in all these themes. Discussions about these connections can be found in a series of papers~\cite{Lin:2021hqs,Lin:2022aqf,Lin:2022flo,Lin:2022agc,Lin:2023orb,Kudler-Flam:2019oru,Rolph:2021nan}. However, our main point is that, these thread configurations can be endowed with a clear interpretation of a class of quantum states. From now on, we will systematically refer to them as coarse-grained states~\footnote{In previous literature, we often referred to them as distilled states, given their close connection with entanglement distillation tensor networks~\cite{Lin:2020yzf}.}. As the name suggests, coarse-grained states are expected to only characterize the quantum entanglement of holographic quantum systems at a coarse-grained level. The true entanglement structure of a holographic quantum system is expected to be much more complex. The key point is that the study of these coarse-grained states will lead us to recognize the necessity of perfect tensor entanglement in holographic quantum systems. A preliminary discussion can be found in~\cite{Lin:2023orb}. 

In this paper, by using these coarse-grained states to further characterize the some geometric duals of quantum information theory quantities, such as the RT surface corresponding to the entanglement entropy of disconnected regions and the entanglement wedge cross-section (EWCS) corresponding to the entanglement of purification (EoP)~\cite{Nguyen:2017yqw,Takayanagi:2017knl}, we find that even at the coarse-grained level, we encounter unavoidable difficulties. Our core idea is that the introduction of perfect tensor entanglement~\cite{Pastawski:2015qua} naturally resolves these issues.  The introduction of perfect entanglement not only achieves the characterization of the entanglement entropies of disconnected regions~\cite{Lin:2023orb}, but also naturally gives rise to the equivalence between two quantum information theory quantities related to the entanglement wedge cross-section—the entanglement of purification~\cite{Nguyen:2017yqw,Takayanagi:2017knl} and the balanced partial entropy (BPE)~\cite{Wen:2021qgx,Camargo:2022mme,Wen:2022jxr}. In other words, it allows the so-called BPE to reasonably characterize the inherent correlation in a bipartite mixed state. Moreover, to demonstrate our ideas more clearly and concretely, we construct a thread configuration with a Multi-Scale Entanglement Renormalization Ansatz (MERA) structure~\cite{Vidal:2007hda,Vidal:2008zz,Vidal:2015,Swingle:2009bg,Swingle:2012wq}, and associate it with a coarse-grained state characterized by perfect tensor entanglement. We further explore the connection between the MERA structure and kinematic space~\cite{Czech:2015kbp,Czech:2015qta}. The construction also has an inspiring role in understanding the relationship between MERA tensor networks and bit threads~\cite{Chen:2018ywy}.

The structure of this paper is as follows: In Section 2, we provide a review of refined thread configurations, coarse-grained states, and the thread-state correspondence. In Section 3, motivated by the thread perspective for characterizing the entanglement entropy of disconnected regions and the entanglement wedge cross-section, we propose the necessity of introducing perfect tensor entanglement. In Section 4, we construct a thread configuration with the MERA structure and its corresponding coarse-grained state to illustrate our proposal. This thread configuration provides an example of thread-state correspondence and is closely related to kinematic space. In Sections 5 and 6, we interconnect all intersecting threads in the MERA-structure thread configuration to introduce the perfect tensor entanglement, and demonstrate that it can nicely solve the problems proposed in Section 3. Section 7 concludes the paper with discussions.


\section{Background Review}\label{sec2}

\begin{figure}
     \centering
     \begin{subfigure}[b]{0.4\textwidth}
         \centering
         \includegraphics[width=\textwidth]{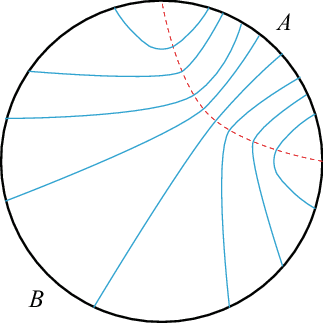}
         \caption{}
         \label{2.1.1a}
     \end{subfigure}
     \hfill
     \begin{subfigure}[b]{0.4\textwidth}
         \centering
         \includegraphics[width=\textwidth]{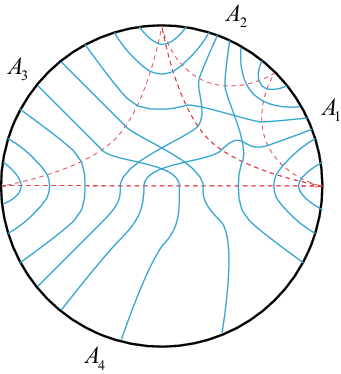}
         \caption{}
         \label{2.1.1b}
     \end{subfigure}
     \hfill
     \begin{subfigure}[b]{1\textwidth}
         \centering
         \includegraphics[width=\textwidth]{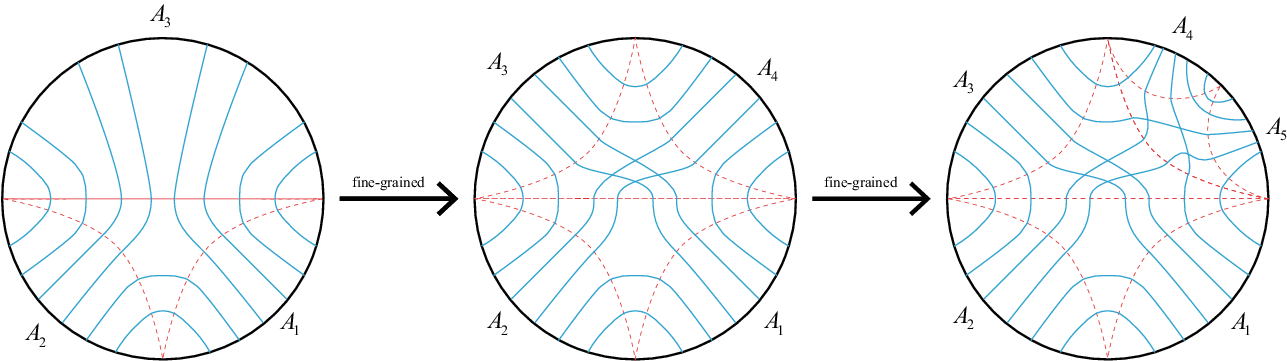}
         \caption{}
         \label{refine}
     \end{subfigure}
     \caption{(a) A ``thread'' picture characterizing the entanglement entropy between two complementary regions. (b) A more refined thread configuration characterizing a set of entanglement entropies involving more subregions. (c) By iteratively dividing the quantum system, more and more refined thread configurations can be constructed, characterizing the entanglement structure at more and more refined levels, see details in~\cite{Lin:2022flo}.  Here the threads are schematically represented as blue lines, and the RT surfaces are represented as red dashed lines.}
\end{figure}

\subsection{Refined Thread Configurations}\label{sec21}

A naive illustration of the entanglement structure revealed by the RT formula is shown in Figure \ref{2.1.1a}. Consider the entanglement entropy of a subsystem $A$ and its complement $B$ in a pure state of a CFT dual to a pure AdS space. We can envision a family of uninterrupted threads connecting $A$ and its complement $B$, passing through the RT surface $\gamma_A$. The number of these threads, denoted as $N_{AB}$, is exactly equal to the entanglement entropy $S(A)$ between $A$ and $B$:
\be\label{ab} N_{AB} = S_A.\ee
These threads are commonly referred to as ``bit threads"~\cite{Freedman:2016zud,Cui:2018dyq,Headrick:2017ucz,Headrick:2022nbe}.

We can do better. We can construct a series of increasingly refined thread configurations that can be used to calculate the entanglement entropies for more than one region~\cite{Headrick:2020gyq,Lin:2020yzf,Lin:2021hqs,Lin:2022flo,Lin:2022agc}. As shown in Figure \ref{2.1.1b}, we can further decompose region $A$ into $A = A_1 \cup A_2$, and $B$ into $B = A_3 \cup A_4$. Thus, we can construct a more refined thread configuration that can calculate the entanglement entropies between six connected regions and their complements, including: $A_1$, $A_2$, $A_3$, $A_4$, $A = A_1 \cup A_2$, and $A_2 \cup A_3$. In other words, the number of threads connecting these six regions and their complements is exactly equal to the corresponding entanglement entropy. Similarly, we can further divide the quantum system $M$ into more adjacent and non-overlapping basic regions $A_1$, $A_2$, ..., $A_N$, and then obtain corresponding more refined thread configurations (see Fig. \ref{refine})~\footnote{These thread configurations are referred to as ``locking" thread configurations in the literature~\cite{Headrick:2020gyq}. Note that here we have carefully drawn the threads to appear perpendicular to the RT surfaces they pass through, in keeping with the conventional property of bit threads. However, in the sense of coarse-grained states, only the topology is really important, and we have not yet seriously considered the exact trajectories of these threads.}. This process can be iterated (as long as each basic region is still much larger than the Planck length to ensure the applicability of the RT formula). Here, we define basic regions such that:
\be A_i \cap A_j = \emptyset, \quad \bigcup A_i = M.\ee

Next, we can define a function $N_{ij} \equiv N_{A_i \leftrightarrow A_j}$ for each pair of basic regions $A_i$ and $A_j$, representing the number of threads connecting $A_i$ and $A_j$. The papers~\cite{Lin:2021hqs,Lin:2022flo,Lin:2022agc} impose the following physical requirement on the thread configuration: the set of threads $\{N_{ij}\}$ should satisfy:
\be\label{equ} {S_{a(a + 1) \ldots b}} = \sum\limits_{i,j} {{N_{ij}}} ,\;\;{\rm{where}}\;i \in \{ a,a + 1, \ldots ,b\} ,\;j \notin \{ a,a + 1, \ldots ,b\} .\ee
Here, $S_{a(a + 1) \ldots b}$ represents the entanglement entropy $S_A$ of a connected composite region $A = A_{a(a + 1) \ldots b} \equiv A_a \cup A_{a + 1} \cup \ldots \cup A_b$. This equation can be intuitively understood as the entanglement entropy between $A$ and its complement $\bar{A}$ coming from the sum of $N_{ij}$ between basic regions $A_i$ within $A$ and basic regions $A_j$ within the complement $\bar{A}$. Thread configurations satisfying condition~(\ref{equ}) are commonly referred to as ``locking" thread configurations, borrowing terminology from network flow theory. Solving condition~(\ref{equ}), the first thing we find is that the number of threads connecting two basic regions is precisely given by the so-called conditional mutual information. In other words, the conditional mutual information characterizes the correlation between two regions $A_i$ and $A_j$ separated by a distance $L$:
\be\label{cmi}N_{A_i \leftrightarrow A_j} = \frac{1}{2} I(A_i, A_j | L) \equiv \frac{1}{2} [S(A_i \cup L) + S(A_j \cup L) - S(A_i \cup L \cup A_j) - S(L)].\ee
Here, $L = A_{(i + 1) \ldots (j - 1)}$ represents the region between $A_i$ and $A_j$, which is a composite region consisting of many basic regions and also represents the distance between $A_i$ and $A_j$.

In fact, this class of refined thread configurations is closely related to various concepts proposed from different perspectives in holographic duality research, including kinematic space~\cite{Czech:2015kbp,Czech:2015qta}, entropy cone~\cite{Bao:2015bfa,Hubeny:2018ijt,Hubeny:2018trv,HernandezCuenca:2019wgh}, and holographic partial entanglement entropy~\cite{Vidal:2014aal,Wen:2019iyq,Wen:2018whg,Wen:2020ech,Han:2019scu,Kudler-Flam:2019oru,Lin:2021hqs}. These connections are, in a sense, natural and easy to obtain, especially where the conditional mutual information plays a central role. For example, it is defined as the volume measure in kinematic space~\cite{Czech:2015kbp,Czech:2015qta}, characterizing the density of entanglement entropy. Discussions about these connections can be found in a series of articles~\cite{Lin:2021hqs,Lin:2022aqf,Lin:2022flo,Lin:2022agc,Lin:2023orb,Kudler-Flam:2019oru,Rolph:2021nan}. The key point is that, in our framework, we will understand these refined thread configurations as a kind of coarse-grained state of the holographic quantum system~\cite{Lin:2022flo,Lin:2022agc,Lin:2023orb}, which only characterizes the entanglement structure of the holographic quantum system at a coarse-grained level. Our approach is to explore the properties that these entanglement structures should have by studying them at this level.

\subsection{Coarse-Grained State}\label{sec22}

Now, we will refer to these thread configurations as representing the coarse-grained state of the holographic quantum system. In~\cite{Lin:2022flo,Lin:2022agc,Lin:2023orb}, this is also referred to as thread-state correspondence. In simple terms, the idea is that in such a locking thread configuration, each thread can be understood as a pair of maximally entangled qudits. One end of the thread corresponds to one qudit. For example, let's take $d=2$, so one end of the thread corresponds to a qubit. Thus, a thread corresponds to
\be\label{qud}\left| \text{thread} \right\rangle = \frac{1}{{\sqrt{2}}} \left( \left| 00 \right\rangle + \left| 11 \right\rangle \right).\ee
Then the direct product of the states of all threads in the thread configuration gives a coarse-grained state of the quantum system.
\be{\left| \Psi \right\rangle_{\text{coarse}}} = \prod_{\text{all thread}} {\left| \text{thread} \right\rangle}.\ee
~\cite{Lin:2022flo,Lin:2022agc,Lin:2023orb} argues that if we take the partial trace of the coarse-grained state ${\left| \Psi \right\rangle_{\text{coarse}}}$ to obtain reduced density matrices for various connected regions $\{A_i, A_iA_{i + 1}, A_iA_{i + 1}A_{i + 2}, \ldots\}$ and calculate the corresponding von Neumann entropies, we exactly obtain a set of correct holographic entanglement entropies.

In fact, the idea of thread-state correspondece expresses more than what the expression (\ref{qud}) presents. By combining with the surface/state duality~\cite{Miyaji:2015yva,Miyaji:2015fia} or tensor network models, each thread actually represents not only the entanglement between the two endpoints of the thread but also the entanglement between all degrees of freedom that the thread passes through in the holographic bulk. For this, we can agree on a thread-state rule: each thread is in a state~\cite{Lin:2022flo,Lin:2022agc,Lin:2023orb}
\be\label{qub}\left| \text{thread} \right\rangle  = \frac{1}{{\sqrt 2 }}(\left| \text{red} \right\rangle  + \left| \text{blue} \right\rangle ).\ee
Then, the entanglement between multiple sites (or qudits) strung together by a single thread follows the rules: each $\left| \text{red} \right\rangle$ state actually represents that the qudits the thread passes through are all in their own $\left| 0 \right\rangle $ state, and each $\left| \text{blue} \right\rangle$ state actually represents that the qudits the thread passes through are in their own $\left| 1 \right\rangle $ state, that is,
\be\begin{array}{l}
\left| \text{red} \right\rangle = \left| 0_1 0_2 \cdots 0_n \right\rangle \\
\left| \text{blue} \right\rangle = \left| 1_1 1_2 \cdots 1_n \right\rangle
\end{array}.\ee
In fact, we will see explicit examples of this interpretation in the following sections.



\section{Motivation and Proposal: Necessity of Perfect Entanglement}\label{sec3}

As the name suggests, the coarse-grained state is just a characterization of quantum entanglement at the coarse-grained level of the holographic quantum system. The entanglement structure of a genuine holographic quantum system is expected to be much more complex. The key point is that the study of these coarse-grained states will lead us to discover some properties of the entanglement structure in holographic quantum systems. In particular, in this paper, we will systematically demonstrate how the study of these coarse-grained states will lead to the necessity of perfect tensor entanglement in holographic quantum systems, with our preliminary work available in~\cite{Lin:2023orb}. This section will demonstrate this in two aspects, where Section 3.1 is a restatement of the conclusions of~\cite{Lin:2023orb}, while Section 3.2 is entirely new. In Section 4, we will support our argument by constructing a more interesting locking thread configuration.

\subsection{Characterizing Entanglement Entropy of Disconnected Regions}\label{sec31}

\begin{figure}
    \centering
    \includegraphics[scale=0.7]{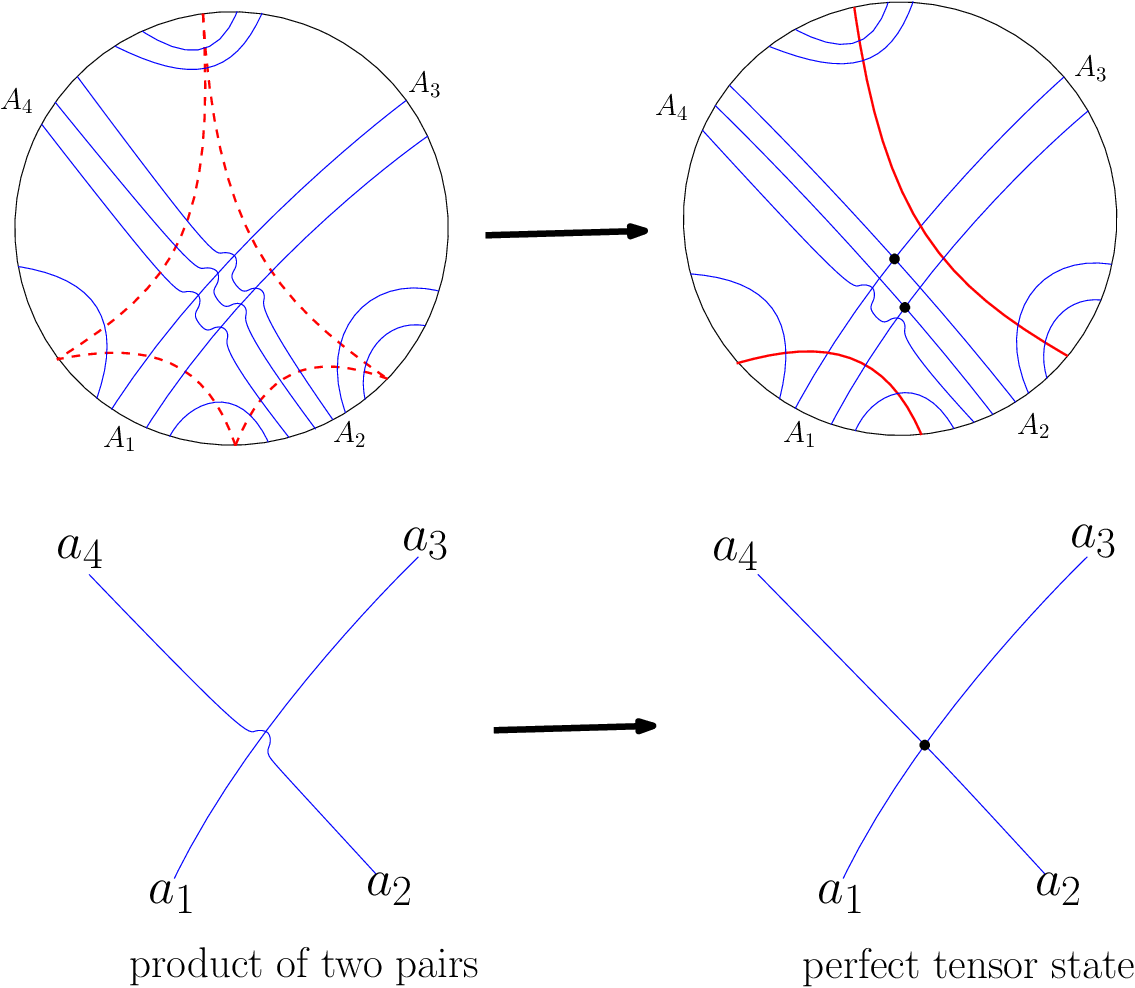}
    \caption{The thread configuration in the upper left corresponds to a coarse-grained state constructed from the direct product of all pairwise entangled states, from which the entanglement entropies of all the connected subregions involved can be calculated. The top-right figure shows that in order for the original coarse-grained state to further characterize the entanglement entropy of an disconnected region ${A_1} \cup {A_3}$, perfect entanglement must be introduced. The key point is illustrated at the bottom. Consider a thread connecting a site $a_1$ in $A_1$ and a site $a_3$ in $A_3$, and couple it to a thread connecting a site $a_2$ in $A_2$ and a site $a_4$ in $A_4$, so that the overall state of four qudits changes from (\ref{ori}) to (\ref{per}).}
    \label{fig3.1.1}
\end{figure}

Characterizing the holographic entanglement entropy of disconnected regions using the coarse-grained state is non-trivial. Without loss of generality, let's focus on the entanglement entropy ${S_R}$ of a non-connected region $R = {A_1} \cup {A_3}$ in the diagram~\ref{fig3.1.1}. Suppose we already have a coarse-grained state ${\left| \Psi \right\rangle _{coarse}}$ corresponding to the top-left of Fig.~\ref{fig3.1.1}, which can characterize the entanglement entropies of all connected regions, i.e., $\{ {A_1},\;{A_2},\;{A_3},\;{A_4},\;{A_1} \cup {A_2},\;{A_2} \cup {A_3}\} $. We can further refine this coarse-grained state to obtain a new coarse-grained state ${\left| {\Psi '} \right\rangle _{coarse}}$, such that the new coarse-grained state ${\left| {\Psi '} \right\rangle _{coarse}}$ can also characterize the entanglement entropy of the disconnected region $R$.

A simple method is pointed out in~\cite{Lin:2023orb}. Note that in the original thread configuration (see the top-left of Fig.~\ref{fig3.1.1}), intersecting threads do not influence each other. For example, consider two intersecting threads $a_1a_3$ and $a_2a_4$, apply the thread-state correspondence, and take $d=3$. Then the overall state of the intersecting threads is simply the direct product of two pair-entangled states, i.e.,
\be\label{ori}\left| {{a_1}{a_2}{a_3}{a_4}} \right\rangle  = \frac{1}{{\sqrt 3 }}(\left| {{0_{{a_1}}}{0_{{a_3}}}} \right\rangle  + \left| {{1_{{a_1}}}{1_{{a_3}}}} \right\rangle  + \left| {{2_{{a_1}}}{2_{{a_3}}}} \right\rangle ) \otimes \frac{1}{{\sqrt 3 }}(\left| {{0_{{a_2}}}{0_{{a_4}}}} \right\rangle  + \left| {{1_{{a_2}}}{1_{{a_4}}}} \right\rangle  + \left| {{2_{{a_2}}}{2_{{a_4}}}} \right\rangle ).\ee
Now, the idea is to entangle these two non-interacting threads, making the overall state of the four qutrits $a_1, a_3, a_2, a_4$ a specific entangled state as follows:
\be\label{per}\begin{array}{l}
\left| {{a_1}{a_2}{a_3}{a_4}} \right\rangle  = \frac{1}{3}(\left| {{0_{{a_1}}}{0_{{a_3}}}{0_{{a_2}}}{0_{{a_4}}}} \right\rangle  + \left| {{1_{{a_1}}}{1_{{a_3}}}{1_{{a_2}}}{0_{{a_4}}}} \right\rangle  + \left| {{2_{{a_1}}}{2_{{a_3}}}{2_{{a_2}}}{0_{{a_4}}}} \right\rangle \\
\quad \quad \quad \quad \;\,\,\, + \left| {{0_{{a_1}}}{1_{{a_3}}}{2_{{a_2}}}{1_{{a_4}}}} \right\rangle  + \left| {{1_{{a_1}}}{2_{{a_3}}}{0_{{a_2}}}{1_{{a_4}}}} \right\rangle  + \left| {{2_{{a_1}}}{0_{{a_3}}}{1_{{a_2}}}{1_{{a_4}}}} \right\rangle \\
\quad \quad \quad \quad \,\;\,\, + \left| {{0_{{a_1}}}{2_{{a_3}}}{1_{{a_2}}}{2_{{a_4}}}} \right\rangle  + \left| {{1_{{a_1}}}{0_{{a_3}}}{2_{{a_2}}}{2_{{a_4}}}} \right\rangle  + \left| {{2_{{a_1}}}{1_{{a_3}}}{0_{{a_2}}}{2_{{a_4}}}} \right\rangle )
\end{array}.\ee
The key is that the original state (\ref{ori}) is not symmetric about the four qutrits. Note that in it, the entanglement entropy between qutrits  $a_1 \cup a_2$ with $a_3 \cup a_4$ is $2\log 3$, while the entanglement entropy between qutrits $a_1 \cup a_2$ with $a_2 \cup a_4$ is 0. The reason is simple because $a_1$ and $a_3$ are at the two ends of the same thread, and $a_2$ and $a_4$ are at the two ends of the same thread. And these two threads are direct-product. On the other hand, the new state is symmetric about the four qutrits.
The state (\ref{per}) is commonly referred to as a perfect tensor state in the literature. It can be more compactly written as (we adopt Einstein's index summation convention and omit the summation symbol.)
\be\label{ksi}\left| \chi \right\rangle  = {T_{\alpha \beta \mu \nu }}\left| {\alpha \beta \mu \nu } \right\rangle, \ee
where $\left| {\alpha \beta \mu \nu } \right\rangle  = \left| \alpha \right\rangle  \otimes \left| \beta \right\rangle  \otimes \left| \mu \right\rangle  \otimes \left| \nu \right\rangle $, and
\be{T_{0000}} = {T_{1110}} = {T_{2220}} = {T_{0121}} = {T_{1201}} = {T_{2011}} = {T_{0212}} = {T_{1022}} = {T_{2102}} = \frac{1}{3},\ee
while other components are 0. ${T_{\alpha \beta \mu \nu }}$ is an example of a (rank four) perfect tensor, and correspondingly, $\left| \chi \right\rangle $ is a perfect tensor state. The two ``knotted" threads in the bottom-right of Fig.~\ref{fig3.1.1} form a vertex extending four legs, essentially the graphical representation of ${T_{\alpha \beta \mu \nu }}$ in tensor network language. The key is to realize that this state has an interesting property: for any one of the four qutrits, the entanglement entropy between it and its complement is $\log 3$, and for any two of them with their complement, the entanglement entropy is $2\log 3$.
Perfect tensor states are well-known ingredients in the holographic HaPPY code used for quantum error correction~\cite{Pastawski:2015qua} and are also known as Absolutely Maximally Entangled (AME) states in quantum information theory~\cite{Helwig:2012nha,Helwig:2013qoq}. More generally, perfect tensors can be defined equivalently in two ways:

Definition 1: A $2s$-perfect tensor is a $2s$-qudit pure state for any positive integer $s$ such that the reduced density matrix involving any $s$ qudits is maximally mixed.

Definition 2: A $2s$-perfect tensor is a $2s$-qudit pure state for any positive integer $k \leq s$ such that the mapping from the states of any $k$ qudits to the states of the remaining $2s-k$ qudits is an isometric isomorphism.

In summary, by replacing the tensor product states with perfect tensor states, it is possible to characterize the entanglement entropy of disconnected regions using coarse-grained states. More detailed discussions can be found in~\cite{Lin:2023orb}. We will further indicate in Section \ref{sec51} that perfect tensor entanglement is inevitable for characterizing the entanglement entropies of disconnected regions.

\subsection{Characterizing the Entanglement Wedge Cross Section}\label{sec32}

In this subsection, we point out another noteworthy phenomenon that once again indicates the inevitability of perfect tensor entanglement for the entanglement structure of holographic quantum systems at the coarse-grained level. This phenomenon is related to the holographic entanglement wedge cross section.

\begin{figure}
     \centering
     \begin{subfigure}[b]{0.4\textwidth}
         \centering
         \includegraphics[width=\textwidth]{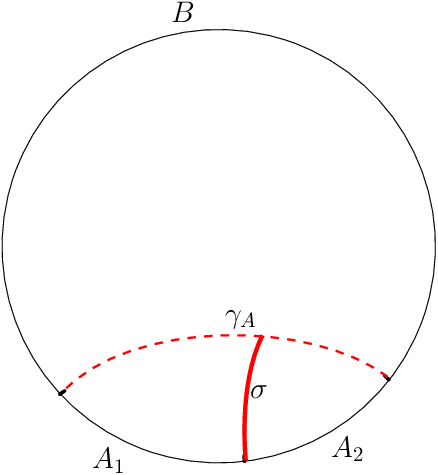}
         \caption{}
         \label{fig3.2.1a}
     \end{subfigure}
     \hfill
     \begin{subfigure}[b]{0.4\textwidth}
         \centering
         \includegraphics[width=\textwidth]{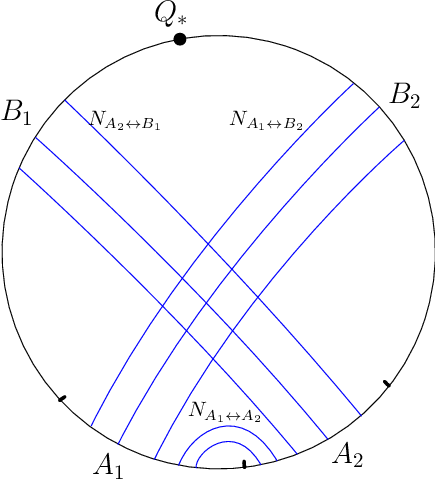}
         \caption{}
         \label{fig3.2.1b}
     \end{subfigure}
     \caption{(a)	The correlation between two adjacent subregions $A_1$ and $A_2$ can be holographicly measured by the area of the entanglement wedge cross section ${\sigma _{{A_1}:{A_2}}}$, represented by the red solid line. (b) The BPE method to calculate the area of ${\sigma _{{A_1}:{A_2}}}$: divide the complement $B$ into two parts such that ${N_{{A_1} \leftrightarrow {B_2}}} = {N_{{A_2} \leftrightarrow {B_1}}}$.}
\end{figure}

Taking Fig.~\ref{fig3.2.1a} as an example, let's consider the correlation between two adjacent subregions $A_1$ and $A_2$ in the CFT. Within the framework of holographic principles, this quantity can be measured by the area of a so-called entanglement wedge cross section ${\sigma _{{A_1}:{A_2}}}$~\cite{Nguyen:2017yqw,Takayanagi:2017knl}. The definition of the entanglement wedge cross section ${\sigma _{{A_1}:{A_2}}}$ is as follows: first, define the entanglement wedge $W(A)$ of $A = {A_1} \cup {A_2}$ as the bulk region enclosed by $A$ and its corresponding RT surface ${\gamma _A}$. Then, a minimal extremal surface ${\sigma _{{A_1}:{A_2}}}$ can be defined, satisfying: 1, it divides the entanglement wedge $W(A)$ into two parts, one entirely touching ${A_1}$ and the other entirely touching ${A_2}$. 2, among all extremal surfaces satisfying condition 1, select the one with the smallest area.

Now, there is a holographic method to calculate the area of this surface~\cite{Wen:2021qgx}. As shown in Fig.~\ref{fig3.2.1b}, first, we find a point $Q_*$ on the complement $B$ of $A$ to divide $B$ into two parts $B_1$ and $B_2$. Consequently, we can construct a locking thread configuration corresponding to the basic region selection $\left\{ {{A_1},\;{A_2},\;{B_1},\;{B_2}} \right\}$ to characterize the entanglement structure at this coarse-grained level. The requirement is to find a point ${Q_*}$ such that in its corresponding thread configuration, the number of threads connecting $A_1$ and $B_2$ is equal to the number of threads connecting $A_2$ and $B_1$:
\be{N_{{A_1} \leftrightarrow {B_2}}} = {N_{{A_2} \leftrightarrow {B_1}}}.\ee
Then, we can obtain
\be\label{sig}\frac{{\text{{Area}}\left( {{\sigma _{{A_1}:{A_2}}}} \right)}}{{4{G_N}}} = {N_{{A_1} \leftrightarrow {B_2}}} + {N_{{A_1} \leftrightarrow {A_2}}}.\ee
This interesting fact was first proposed in the language of partial entanglement entropy in the paper~\cite{Wen:2021qgx} and is known as the Balanced Partial Entanglement (BPE) method. The right-hand side of (\ref{sig}) is named as BPE.

So, how do we understand this method of probing the geometric area? Essentially, the area of the surface ${\sigma _{{A_1}:{A_2}}}$ measures the correlation between the two parts ${A_1}$ and ${A_2}$ in $A$\footnote{In previous studies, this correlation has been understood as various quantum information theory quantities such as entanglement of purification (EoP)~\cite{Nguyen:2017yqw,Takayanagi:2017knl}, balanced partial entanglement (BPE)~\cite{Wen:2021qgx,Camargo:2022mme,Wen:2022jxr}, reflected entropy~\cite{Dutta:2019gen}, logarithmic negativity~\cite{Kudler-Flam:2018qjo,Kusuki:2019zsp}, odd entropy~\cite{Tamaoka:2018ned}, differential purification~\cite{Espindola:2018ozt}, etc.}. When we apply the concept of coarse-grained states of thread configurations to understand this ``experimental fact," we will once again see the necessity of perfect tensor states.

The key point remains that we must ``entangle" the threads connecting $A_1$ and $B_2$ with the threads connecting $A_2$ to $B_2$ to form a perfect tensor state (\ref{per}) about the four qutrits. As shown in Fig.~, once we handle the coarse-grained state in this way, it becomes clear that the correlation between $A_1$ and $A_2$ is precisely composed of two parts. One part (the second term in (\ref{sig})) is contributed by the bipartite entanglement
\be\label{a1a2}\left| {{a_1}{a_2}} \right\rangle  = \frac{1}{{\sqrt 3 }}(\left| {{0_{{a_1}}}{0_{{a_2}}}} \right\rangle  + \left| {{1_{{a_1}}}{1_{{a_2}}}} \right\rangle  + \left| {{2_{{a_1}}}{2_{{a_2}}}} \right\rangle ),\ee
where between $a_1$ and $a_2$ there exists $\log 3$ of entanglement. The other part (the first term in (\ref{sig})) is contributed by the perfect tensor entanglement
\be\label{a1b2}\begin{array}{l}
\left| {{a_1}{b_2}{a_2}{b_1}} \right\rangle  = \frac{1}{3}(\left| {{0_{{a_1}}}{0_{{b_2}}}{0_{{a_2}}}{0_{{b_1}}}} \right\rangle  + \left| {{1_{{a_1}}}{1_{{b_2}}}{1_{{a_2}}}{0_{{b_1}}}} \right\rangle  + \left| {{2_{{a_1}}}{2_{{b_2}}}{2_{{a_2}}}{0_{{b_1}}}} \right\rangle \\
\quad \quad \quad \quad \;\,\,\, + \left| {{0_{{a_1}}}{1_{{b_2}}}{2_{{a_2}}}{1_{{b_1}}}} \right\rangle  + \left| {{1_{{a_1}}}{2_{{b_2}}}{0_{{a_2}}}{1_{{b_1}}}} \right\rangle  + \left| {{2_{{a_1}}}{0_{{b_2}}}{1_{{a_2}}}{1_{{b_1}}}} \right\rangle \\
\quad \quad \quad \quad \,\;\,\, + \left| {{0_{{a_1}}}{2_{{b_2}}}{1_{{a_2}}}{2_{{b_1}}}} \right\rangle  + \left| {{1_{{a_1}}}{0_{{b_2}}}{2_{{a_2}}}{2_{{b_1}}}} \right\rangle  + \left| {{2_{{a_1}}}{1_{{b_2}}}{0_{{a_2}}}{2_{{b_1}}}} \right\rangle )
\end{array},\ee  
where $a_1$ is symmetrically entangled with the other three qutrits in its complement, which introduces $\log 3$ of correlation between $a_1$ and $a_2$.

Once again, we see the necessity of perfect state entanglement. Because if we only use bipartite entanglement, that is, if we do not ``entangle" the threads connecting $A_1$ and $B_2$ with the threads connecting $A_2$ to $B_2$, there will be no correlation between $a_1$ and $a_2$ at this point, as seen in its corresponding expression:
\be\left| {{a_1}{b_2}{a_2}{b_1}} \right\rangle  = \frac{1}{{\sqrt 3 }}(\left| {{0_{{a_1}}}{0_{{b_2}}}} \right\rangle  + \left| {{1_{{a_1}}}{1_{{b_2}}}} \right\rangle  + \left| {{2_{{a_1}}}{2_{{b_2}}}} \right\rangle ) \otimes \frac{1}{{\sqrt 3 }}(\left| {{0_{{a_2}}}{0_{{b_1}}}} \right\rangle  + \left| {{1_{{a_2}}}{1_{{b_1}}}} \right\rangle  + \left| {{2_{{a_2}}}{2_{{b_1}}}} \right\rangle ).\ee
Because at this point, the two threads $a_1b_2$ and $a_2b_1$ are independent of each other. In this way, we would miss the first contribution in (\ref{sig}) and fall into contradiction.

In this paper, we will only consider the case where $A_1$ and $A_2$ are adjacent. Similar results, as in (\ref{sig}), will still appear for the case where $A_1$ and $A_2$ are not adjacent. However, the physical analysis is entirely similar.


\section{A More Concrete Model: Thread Configurations with MERA Structure}

In this section, we will construct a more nontrivial thread configuration to further elucidate our proposal introduced in the previous section, namely, the necessity of perfect tensor entanglement for holographic quantum systems at the coarse-grained level. Specifically, we construct a thread configuration with a Multiscale Entanglement Renormalization Ansatz (MERA) structure~\cite{Vidal:2007hda,Vidal:2008zz,Vidal:2015,Swingle:2009bg,Swingle:2012wq}. MERA structure originated as a tensor network method invented for characterizing the ground state of critical systems that do not satisfy the area law~\cite{Vidal:2007hda,Vidal:2008zz,Vidal:2015}. It has been considered in~\cite{Swingle:2009bg,Swingle:2012wq,Milsted:2018san} to simulate a time slice of AdS spacetime in the holographic duality and was later proposed to be closely related to the kinematic space corresponding to AdS space in~\cite{Czech:2015kbp,Czech:2015qta}. Therefore, it is a highly important and insightful structure in the research of holographic duality.

Our approach is as follows: first, we will show that the MERA tensor network automatically generates a locking thread configuration. In other words, we construct a specific thread configuration with a MERA structure. This allows us to study its corresponding coarse-grained state. Consequently, we can further introduce perfect tensor entanglement to this coarse-grained state and demonstrate how this procedure resolves issues encountered in characterizing the holographic entanglement entropies of disconnected regions and the correlations dual to entanglement wedge cross sections.

\begin{figure}
    \centering
    \includegraphics[scale=0.7]{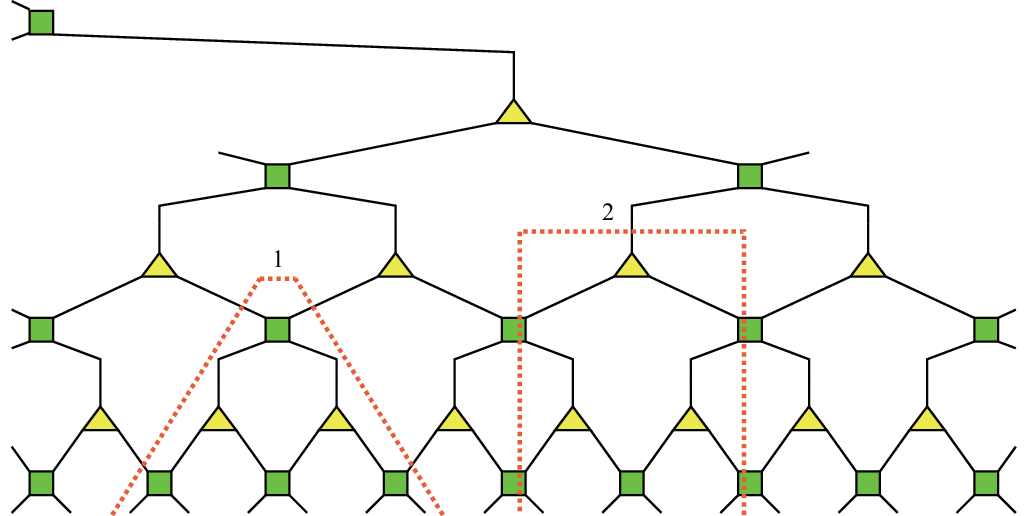}
    \caption{The MERA tensor network, composed of two basic tensors. The coarse-grainers are represented by three-leg triangles, the disentangers are represented by four-leg squares. The red dashed lines 1 and 2 represent two choices of defining an RT surface.}
    \label{mera}
\end{figure}

\subsection{MERA Tensor Network Naturally Generates a Thread Configuration}

As shown in Fig.~\ref{mera}, the MERA tensor network is composed of two basic tensors~\cite{Vidal:2007hda,Vidal:2008zz,Vidal:2015}. The first type of tensor is called the coarse-grainer $w$, which is an isometry, satisfying
\be\label{w}{w^\dag }w = 1.\ee  
It is generally represented by a triangle in the diagram and has three legs, indicating that it represents a rank-3 tensor. The other type of tensor is called the disentangers $u$, which is a unitary operator, satisfying
\be\label{u}{u^\dag }u = 1,\;u{u^\dag } = 1.\ee
It is typically represented by a square in the diagram and has four legs, indicating that it represents a rank-4 tensor. MERA is a hierarchical array, representing entanglement at different length scales through alternating layers of coarse-grainers and disentanglers. As the coarse-graining process proceeds, the disentangling layer is responsible for reducing the quantum entanglement of the ground state, which was highly entangled originally. Generally, when the wave function can be written as a direct-product state (an unentangled state), the network terminates. However, for cases with scale invariance, the tensor network has infinite depth. It must be noted that the MERA tensor network is, in fact, a general method for constructing arbitrary quantum states (not limited to critical quantum systems), and its construction recipe depends on the specific form given to the isometry $w$ and the disentangler $u$.

\cite{Swingle:2009bg,Swingle:2012wq} first proposed that an MERA tensor network can simulate a discretization of a time slice of AdS spacetime, particularly, it can characterize the discretized Ryu-Takayanagi (RT) formula. In Fig.~\ref{mera}, we illustrate the concept of the RT surface in the network. It is a ``cut’’ that divides the entire network into two parts satisfying: 1, one part only touches $A$, and the other part only touches $B$, the complement of $A$; 2, among all cuts satisfying condition 1, select the one with the minimum number of cut legs. In the MERA tensor network, it can be argued that the number of legs the cut passes through exactly gives the entanglement entropy of region $A$ (more precisely, an upper bound on the entanglement entropy). Thus, by defining the number of legs the cut passes through as the area of the cut, we return to the RT formula.

\begin{figure}
     \centering
     \begin{subfigure}[b]{0.8\textwidth}
         \centering
         \includegraphics[width=\textwidth]{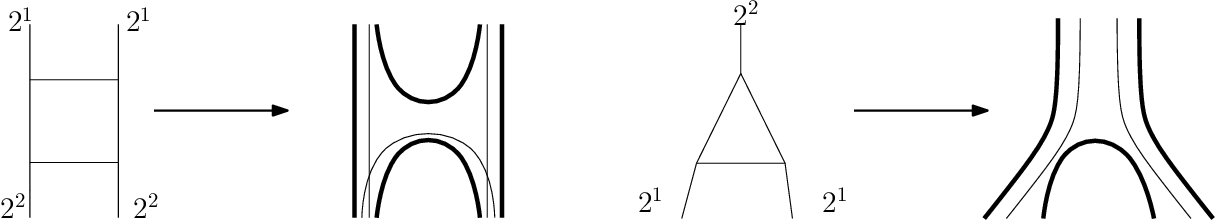}
         \caption{}
         \label{fig4.1.1a}
     \end{subfigure}
     \hfill
     \begin{subfigure}[b]{0.8\textwidth}
         \centering
         \includegraphics[width=\textwidth]{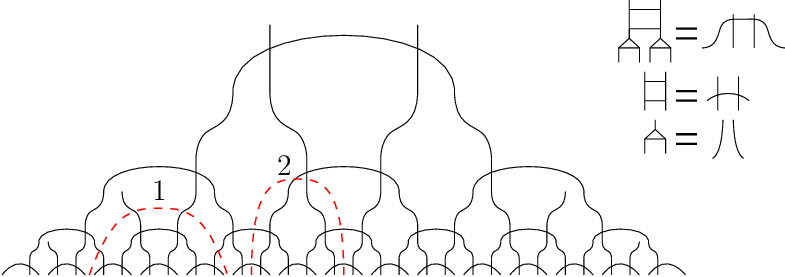}
         \caption{}
         \label{fig4.1.1b}
     \end{subfigure}
     \caption{Regarding the legs of each fundamental tensor as channels carrying thread flows will generate an overall thread configuration. The red dashed lines represent the RT surfaces in this context.
Note that in (a) we choose very simple thread representations of disentanglers and coarse-grainers.}
\end{figure}

Although not necessarily related to our main theme, let us point out one intuitive motivation from the perspective of bit-thread representation that leads to the generation of a thread configuration by the MERA tensor network. In the bit-thread representation, a Planck area is usually allowed to accommodate the passage of one bit thread~\cite{Freedman:2016zud,Cui:2018dyq,Headrick:2017ucz,Headrick:2022nbe}. In other words, the number of bit threads is proportional to the area measure of the channel cross-section. Now, since in the MERA tensor network, the number of legs cut by the RT cut plays the role of the area measure, we can naturally regard each leg as a channel carrying a network flow. We agree that the number of threads passing through this channel is precisely equal to the logarithm of the dimension of the leg. The practical implementation of this idea is shown in Fig.~\ref{fig4.1.1a}. Note that we do not require a general construction; in fact, we choose very simple (even trivial in their own right) representations of disentanglers and coarse-grainers, and such choices should not be expected to capture the ground state of a true conformal field theory (CFT). However, our goal is to construct a coarse-grained state, and what we want to retain is the skeletal structure of MERA itself~\footnote{It was actually proposed in~\cite{Milsted:2018san} that the skeletal structure of the MERA tensor network plays a role in gluing spacetime fragments, which themselves are represented by the so-called euclideon tensors, into an AdS time slice.}. 

Regarding each fundamental tensor as corresponding to channels carrying a subset of thread flows immediately generates an overall thread configuration, as shown in Fig.~\ref{fig4.1.1a}. This configuration can be understood as being composed of many small thread configurations glued together according to the MERA structure. The first thing we can immediately check about this overall thread configuration is that it automatically satisfies the ``locking" property of the thread configurations reviewed in Section~\ref{sec21}. That is, we can directly and precisely calculate the entanglement entropy of a specified connected boundary subregion from this thread configuration. Specifically, consider a boundary subregion $A$ of this thread configuration, which consists of endpoints of a set of threads. Now, if we examine the RT surface $\gamma_A$ corresponding to $A$ in the MERA tensor network, we find that the RT surface exactly divides this thread configuration into two halves, such that each thread passes through $\gamma_A$ at most once, and the number of threads passing through $\gamma_A$ exactly gives the entanglement entropy of $A$ (or precisely the area of $\gamma_A$). Clearly, the red thread 1 marked in the figure as the RT surface is consistent with the interpretation in the thread configuration viewpoint. We can also define the area of the RT surface by counting the number of cut disentanlers~\cite{Nozaki:2013wia}, as shown by the red thread 2, which cuts the ``inner threads’’ of the disentanglers. Nevertheless, cutting the legs of the MERA tensor network can then be interpreted in our construction as cutting the number of threads connecting the interior of region $A$ and the remaining part, and the number of these threads gives the entanglement entropy of $A$, very similar to the picture proposed in the bit-thread representation. Early discussions on the connection between bit-threads and MERA tensor networks can be found in~\cite{Chen:2018ywy}.

\begin{figure}
     \centering
     \begin{subfigure}[b]{0.8\textwidth}
         \centering
         \includegraphics[width=\textwidth]{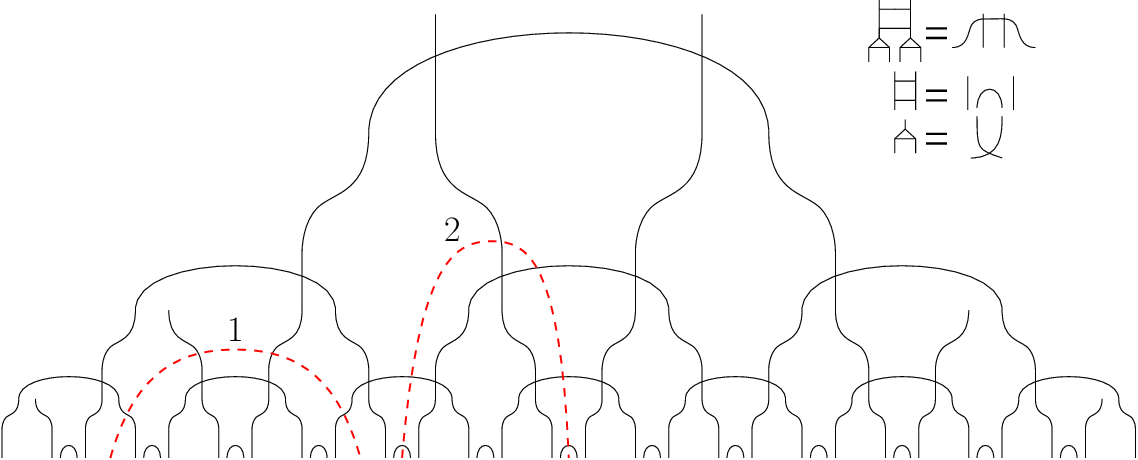}
         \caption{}
         \label{fig4.1.2}
     \end{subfigure}
     \hfill
     \begin{subfigure}[b]{0.8\textwidth}
         \centering
         \includegraphics[width=\textwidth]{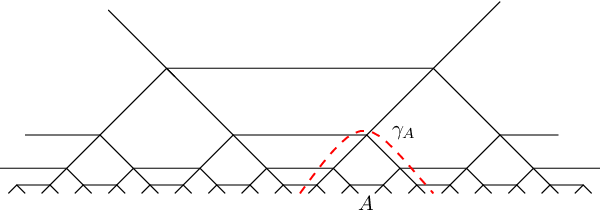}
         \caption{}
         \label{fig4.1.3}
     \end{subfigure}
     \caption{(a) Another thread construction of disentanglers and coarse-grainers lead to a similar thread configuration.  (b) Another representation of the thread configuration with a MERA structure, which has the same topology.}
\end{figure}

Let us point out that if, as shown in Fig.~\ref{fig4.1.2}, we choose another way to construct disentanglers and coarse-grainers, we can also obtain a thread configuration. Moreover, if we examine the trajectory of the RT surface, the above considerations still hold, meaning that we still get a locking thread configuration. In fact, the structures of these two configurations are essentially the same, except for a small difference in the first layer. They are topologically identical and can be equivalently characterized by the diagram in Fig.~\ref{fig4.1.3}. This is because, as shown in Fig.~\ref{fig4.1.2}, taking two coarse-grainers connected by one disentangler in the previous layer as a basic unit, the locally thread configurations obtained by the two construction methods are the same. For symmetry considerations in the diagram, we uniformly define the RT surface as shown in the figure, presenting a shape similar to a light cone. Note that for our current purposes, this diagram has only topological significance; however, we present it as a regular pattern, which, in appearance, is arranged in alternating tiles of quadrilaterals and pentagons. Still, keep in mind that, in the current context, these threads only overlap with each other and are not ``coupled" together (we will discuss the significance of coupling later). This regular pattern facilitates our understanding of the precise correspondence between the MERA structure and kinematic space.

\begin{figure}
    \centering
    \includegraphics{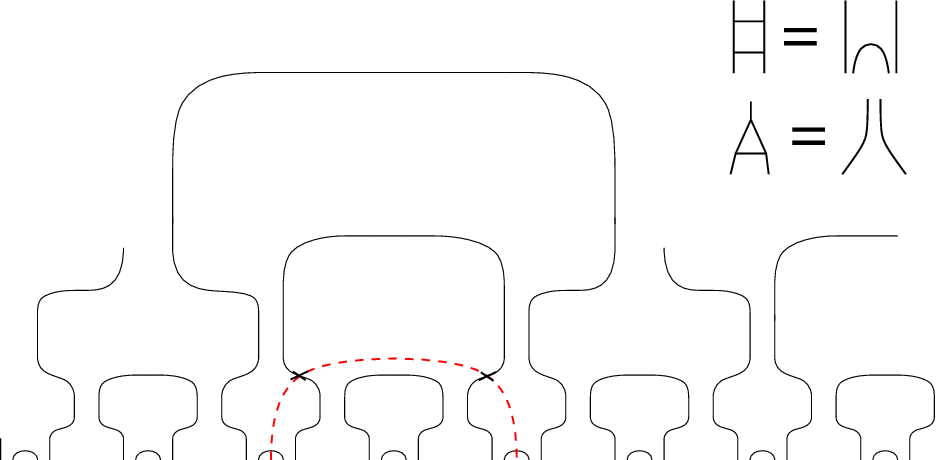}
    \caption{An illegal thread construction for our purposes.}
    \label{fig4.1.4}
\end{figure}
Why do we not choose a thread configuration representation of basic tensors as shown in Fig.~\ref{fig4.1.4}? This is because, in this representation, long-range entanglement cannot be conveniently characterized, as shown in Fig.~\ref{fig4.1.4}. Moreover, the application of the RT formula may also encounter difficulties because the RT surface defined by the minimum cut may pass through the same thread twice. In other words, the MERA state constructed in this way, with the minimum cut, only characterizes the upper bound of the entanglement entropy of the corresponding region and is not exactly equal. Therefore, we choose to construct the thread configuration that satisfies our motivation. In fact, we realize that there are multiple constructions of thread configurations that meet our requirements. For example, as long as we arrange the dimensions of bonds of disentanglers and coarse-grainers appropriately, we can construct more general thread configuration states. However, we only want to discuss what interesting properties the MERA structure will bring to the thread configuration, so we choose the simplest case.

\subsection{Coarse-Grained State Corresponding to the Thread Configuration}
\begin{figure}
     \centering
     \begin{subfigure}[b]{0.8\textwidth}
         \centering
         \includegraphics[width=\textwidth]{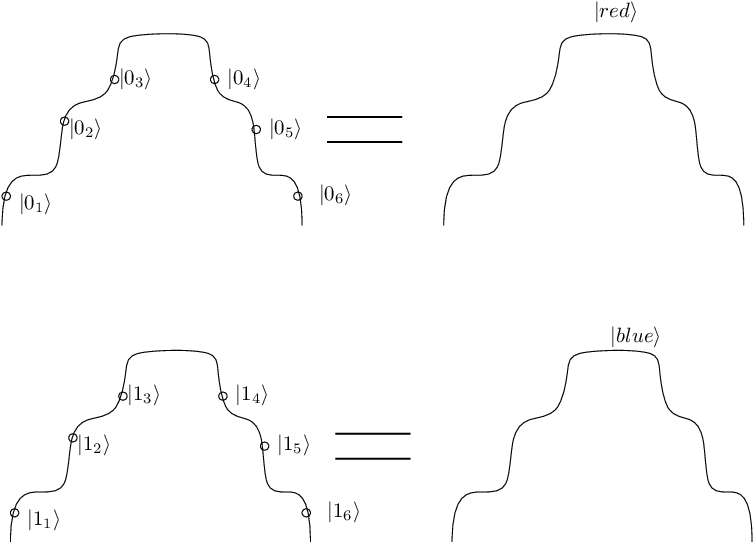}
         \caption{}
         \label{fig4.2.1}
     \end{subfigure}
     \hfill
     \begin{subfigure}[b]{0.8\textwidth}
         \centering
         \includegraphics[width=\textwidth]{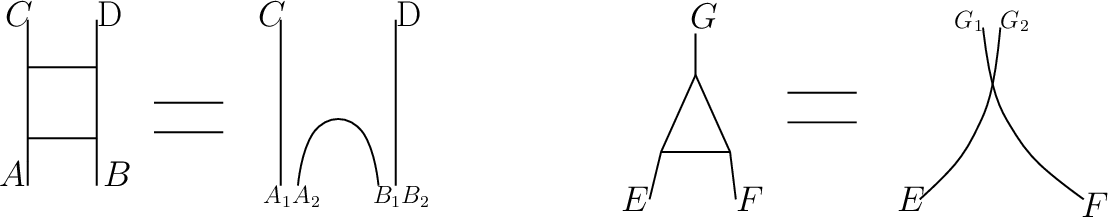}
         \caption{}
         \label{fig4.2.2}
     \end{subfigure}
     \caption{(a) A global thread defined from gluing together the legs of different basic tensors conforms to the rules of the thread-state correspondence. (b)	Essentially, we construct the basic tensors as the coarse-grained states of local thread configurations.}
\end{figure}

Now let's provide a more physical interpretation for the above intuition. In fact, Fig.~\ref{fig4.2.1} perfectly conforms to the rules of the thread-state correspondence reviewed in Section \ref{sec22}. More specifically, each thread can be considered to correspond to a Bell state:
\be\begin{array}{l}
\left| {{\rm{red}}} \right\rangle  = \left| {{0_1}{0_2} \cdots {0_n}} \right\rangle \\
\left| {{\rm{blue}}} \right\rangle  = \left| {{1_1}{1_2} \cdots {1_n}} \right\rangle 
\end{array}.\ee
In this context, that a thread is said to pass through multiple qudits is because, as constructed in Fig.~\ref{fig4.2.1}, a global thread actually results from gluing together the legs of different positions of disentanglers and coarse-grainers. As a result, by focusing on the thread things, we are visualizing the entanglement between different fundamental tensors.

To explicitly show that the thread configurations with MERA structure we constructed can be included in the framework of the thread-state interpretation~\cite{Lin:2022flo,Lin:2022agc,Lin:2023orb}, we provide a physical interpretation for the thread representation of disentanglers and coarse-grainers as shown in Fig.~\ref{fig4.2.2}. Essentially, we construct a special disentangler tensor by using local thread configurations, where each local thread segment represents a Bell pair, and the resulting disentangler is simply the tensor product of these Bell states, i.e.,
\be\label{uth}\left| {u_{{A_1}{A_2}{B_1}{B_2}CD}} \right\rangle = \frac{1}{{\sqrt 2 }}(\left| {0_{{A_1}}0_C} \right\rangle + \left| {1_{{A_1}}1_C} \right\rangle ) \otimes \frac{1}{{\sqrt 2 }}(\left| {0_{{B_2}}0_D} \right\rangle + \left| {1_{{B_2}}1_D} \right\rangle ) \otimes \frac{1}{{\sqrt 2 }}(\left| {0_{{A_2}}0_{{B_1}}} \right\rangle + \left| {1_{{A_2}}1_{{B_1}}} \right\rangle ).\ee
To show that this satisfies the unitary condition~(\ref{u}), we redefine
\be\begin{array}{l}
{\left| {0'0'} \right\rangle _{AB}} = \left| {{0_{{A_1}}}{0_{{B_2}}}} \right\rangle  \otimes \frac{1}{{\sqrt 2 }}(\left| {{0_{{A_2}}}{0_{{B_1}}}} \right\rangle  + \left| {{1_{{A_2}}}{1_{{B_1}}}} \right\rangle )\\
{\left| {0'1'} \right\rangle _{AB}} = \left| {{0_{{A_1}}}{1_{{B_2}}}} \right\rangle  \otimes \frac{1}{{\sqrt 2 }}(\left| {{0_{{A_2}}}{0_{{B_1}}}} \right\rangle  + \left| {{1_{{A_2}}}{1_{{B_1}}}} \right\rangle )\\
{\left| {1'0'} \right\rangle _{AB}} = \left| {{1_{{A_1}}}{0_{{B_2}}}} \right\rangle  \otimes \frac{1}{{\sqrt 2 }}(\left| {{0_{{A_2}}}{0_{{B_1}}}} \right\rangle  + \left| {{1_{{A_2}}}{1_{{B_1}}}} \right\rangle )\\
{\left| {1'1'} \right\rangle _{AB}} = \left| {{1_{{A_1}}}{1_{{B_2}}}} \right\rangle  \otimes \frac{1}{{\sqrt 2 }}(\left| {{0_{{A_2}}}{0_{{B_1}}}} \right\rangle  + \left| {{1_{{A_2}}}{1_{{B_1}}}} \right\rangle )
\end{array}.\ee
With this, the disentangler tensor can be expressed as ${u_{ABCD}}$, and it satisfies
\be{u_{0'0'00}} = {u_{0'1'01}} = {u_{1'0'10}} = {u_{1'1'11}} = \frac{1}{2},\ee
which indeed conforms to the condition~(\ref{u}) and thus is a unitary tensor. Similarly, it is not difficult to verify that the coarse-graining tensor in Fig.~ is an isometry. Likewise, it is a tensor product of three Bell states, where
\be\label{wth}
\left| {{w_{EF{G_1}{G_2}}}} \right\rangle = \frac{1}{{\sqrt 2 }}(\left| {{0_E}{0_{{G_2}}}} \right\rangle + \left| {{1_E}{1_{{G_2}}}} \right\rangle ) \otimes \frac{1}{{\sqrt 2 }}(\left| {{0_F}{0_{{G_1}}}} \right\rangle + \left| {{1_F}{1_{{G_1}}}} \right\rangle )
.\ee
Define
\be\begin{array}{l}
{\left| {0'} \right\rangle _G} = \left| {{0_{{G_2}}}} \right\rangle  \otimes \left| {{0_{{G_1}}}} \right\rangle \\
{\left| {1'} \right\rangle _G} = \left| {{0_{{G_2}}}} \right\rangle  \otimes \left| {{1_{{G_1}}}} \right\rangle \\
{\left| {2'} \right\rangle _G} = \left| {{1_{{G_2}}}} \right\rangle  \otimes \left| {{0_{{G_1}}}} \right\rangle \\
{\left| {3'} \right\rangle _G} = \left| {{1_{{G_2}}}} \right\rangle  \otimes \left| {{1_{{G_1}}}} \right\rangle 
\end{array}.\ee
With this, the coarse-grainer can be written as ${w_{EFG}}$, and it satisfies
\be{w_{000'}} = {w_{011'}} = {w_{102'}} = {w_{113'}} = \frac{1}{2},\ee
which indeed satisfies condition~(\ref{w}). The same method can also be used to explain the state interpretation corresponding to the construction method in Fig.~\ref{fig4.1.2}.

Now, we focus our attention on the thread-state representations (\ref{uth}) and (\ref{wth}) of the fundamental tensors $u$ and $w$. It can be seen that when connecting (contracting) fundamental tensors at different positions, the global thread only needs to exhibit two overall states $\left| {\text{red}} \right\rangle$ and $\left| {\text{blue}} \right\rangle$, as each local thread segment representing a Bell pair essentially provides a unitary mapping between two sites. A global thread vividly represents the transmission path of information.


\subsection{Connection with Kinematic Space}

We have demonstrated that the constructed thread configuration with MERA structure is precisely a locking thread configurations in the sense of Section \ref{sec2}. As a result, this thread configuration can be used to calculate the entanglement entropy between any connected subregion and its complement, which will precisely equal to the number of connecting threads. Moreover, by employing the thread-state correspondence, one can partially trace over the coarse-grained states corresponding to this thread configuration to prove this result. Additionally, in this framework, the number of threads connecting two regions $A_i$ and $A_j$ precisely gives half the value of the conditional mutual information $I({A_i},{A_j}\,|\, {L)}$, where $L$ represents the region between $A_i$ and $A_j$.

Now, it is interesting to compare the thread configuration with MERA structure with kinematic space~\cite{Czech:2015kbp,Czech:2015qta}. \footnote{The connection between kinematic space and bit threads is also discussed in~\cite{Kudler-Flam:2019oru}.} For $AdS_3$, kinematic space is a two-dimensional dual space. The points in this space, denoted in light-cone coordinates as $(u,v)$, are obtained by mapping each pair of points parametrized by coordinates $u$, $v$ on the one-dimensional time slice of the original $CFT_2$. The essential feature of kinematic space is that its metric (or spatial volume density) is defined by the conditional mutual information. In other words, $\frac{1}{2}I({A_i},{A_j}\,|\, {L)}$ is precisely given by the volume of a diamond-shaped region ${\diamondsuit _{{A_i},{A_j}\,|\, L}}$, defined as the region enclosed by the light rays starting from the endpoints of ${A_i}$ and ${A_j}$ at the boundary of kinematic space:
\be\frac{1}{2}I({A_i},{A_j}\,|\, L) = \text{vol}({\diamondsuit _{{A_i},{A_j}\,|\, L}}).\ee

\begin{figure}
     \centering
     \begin{subfigure}[b]{1.0\textwidth}
         \centering
         \includegraphics[width=\textwidth]{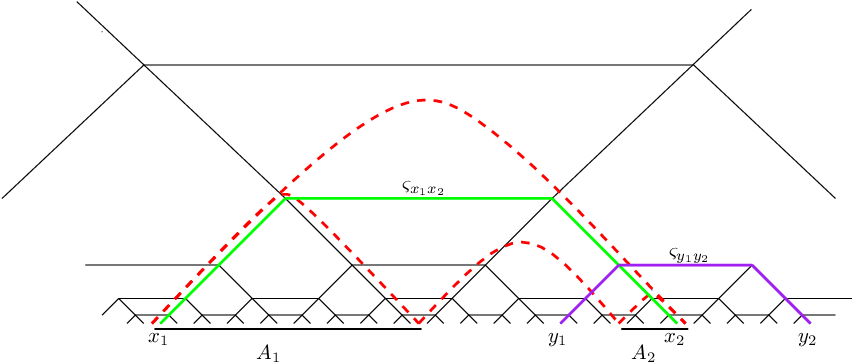}
         \caption{}
         \label{fig5.1.1}
     \end{subfigure}
     \hfill
     \begin{subfigure}[b]{0.8\textwidth}
         \centering
         \includegraphics[width=\textwidth]{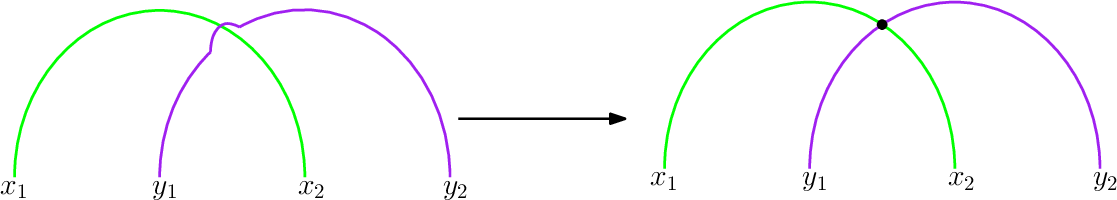}
         \caption{}
         \label{fig5.1.2}
     \end{subfigure}
     \caption{(a) For two nonadjacent subregions (separated by $L$), the green lines here represent the internal entanglement. The number of these internal threads matches the number of disentanglers inside the region ${\diamondsuit _{{A_i},{A_j}\,|\, L}}$ in kinematic space. Furthermore, to characterize the entropy of the union of the two, the green lines should entangle with the purple lines, which represent the entanglement within the complement. Red dashed lines represent RT surfaces. (b) Entangle the threads such that there is $2\log 2$ of entanglement between ${x_1} \cup {x_2}$ and ${y_1} \cup {y_2}$. }
\end{figure}

In~\cite{Czech:2015kbp,Czech:2015qta}, the MERA tensor network is identified as a kinematic space itself, and the conditional mutual information $\frac{1}{2}I({A_i},{A_j}\,|\, L)$ is interpreted as the number of isometries inside the counterpart of ${\diamondsuit _{{A_i},{A_j}\,|\, L}}$ in MERA. Let us carefully examine this viewpoint. In our framework, the volume measure of the kinematic space, represented by the conditional mutual information, is explicitly given by the number of disentangler tensors in MERA. This quantity is also related to the so-called entanglement density in~\cite{Nozaki:2013wia}. The reason is intuitive: as illustrated in Fig.~\ref{fig5.1.1}, considering the RT surfaces ${\gamma _{{A_i}}}$ and ${\gamma _{{A_j}}}$ corresponding to two connected subregions $A_i$ and $A_j$ respectively, $\frac{1}{2}I({A_i},{A_j}\,|\, L)$ represents the number of threads simultaneously crossing ${\gamma _{{A_i}}}$ and ${\gamma _{{A_j}}}$. However, note that now the number of these threads precisely matches the number of ``horizontal lines" inside the counterpart of diamond-shaped region ${\diamondsuit _{{A_i},{A_j}\,|\, L}}$ in our thread configuration. Reviewing the thread representation of disentanglers in Fig.~\ref{fig4.1.1a}, each ``horizontal line" actually corresponds to a disentangler tensor. Therefore, in our framework, $\frac{1}{2}I({A_i},{A_j}\,|\, L)$ precisely corresponds to the number of disentangler tensors inside the diamond-shaped region ${\diamondsuit _{{A_i},{A_j}\,|\, L}}$! The discussion of the connection between entanglement density and conditional mutual information has also been explored in~\cite{Kudler-Flam:2019oru}. It is worth noting that in~\cite{Czech:2015kbp,Czech:2015qta}, the conditional mutual information is argued to be given by the number of isometries (i.e., the coarse grainers). Since in the MERA tensor network, isometries and disentanglers are added successively, the number of both along the light-cone direction is consistent. 

Note that in the general MERA construction, one can only argue that the upper limit of the entanglement entropy of a subregion $A$ is given by the minimum number of cuts. In the model we constructed, however, the entanglement entropy is designed to be strictly saturated, i.e., $S(A)$ is exactly proportional to the minimum number of cuts. Thus, our model helps demonstrate the structural mechanism of the connection between MERA and the nature of kinematic space. Moreover, the equivalence of conditional mutual information and the number of disentangler tensors in our model is exact~\cite{Nozaki:2013wia}. Another thing to note is that, as we have consistently emphasized, what we constructed is not a MERA state that truly characterizes the ground state of a CFT, but rather a coarse-grained state. However, such a MERA state is sufficient to characterize the properties that kinematic space portrays. In a sense, kinematic space does not present many other important aspects of a truly ground-state MERA state characterizing a CFT, but rather reflects its skeletal structure.


\section{ Characterization of Entanglement Entropy for Disonnected Regions}\label{sec5}

\subsection{Perfect Tensor States Couple Bipartite Correlations}\label{sec51}

As reviewed in Section \ref{sec31}, a significant challenge of the coarse-grained state of this thread configuration is its inability to consistently characterize the entanglement entropies of disconnected regions. This paradox is explicitly pointed out in~\cite{Lin:2023orb}. Simply put, in the current scenario, consider a disconnected region $A$ shown in Fig.~\ref{fig5.1.1}, composed of two nonadjacent parts $A_1$ and $A_2$, i.e., $A = {A_1} \cup {A_2}$. The goal is to compute its entanglement entropy. Assuming the sizes of $A_1$ and $A_2$ are relative small compared to the size of their separation, according to the RT formula, then the RT surface should be $\Gamma  = {\gamma _1} \cup {\gamma _2}$, where $\gamma_1$ and $\gamma_2$ are the RT surfaces corresponding to $A_1$ and $A_2$, respectively. However, interpreting the quantum state of thread configurations as a direct product of threads equipped with Bell states cannot correctly provide a self-consistent account of the entanglement entropy for $A = {A_1} \cup {A_2}$. As shown in the figure, the reason is that when the same thread (e.g., the thread connecting the point $x_1$ inside $A_1$ and the point $x_2$ inside $A_2$, denoted as ${\varsigma _{{x_1}{x_2}}}$) simultaneously passes through the surfaces $\gamma_1$ and $\gamma_2$, the entanglement represented by this thread should be regarded as the internal entanglement of $A$, thus not contributing to the entanglement entropy of $A$. More explicitly, calculations based on the direct-product coarse-grained state correctly provides the entanglement entropy for $A_1$ and $A_2$ themselves, but the resulting entanglement entropy for $A$ calculated by the same direct-product coarse-grained state will be less than the expected $S(A) = S({A_1}) + S({A_2})$, and the difference is precisely proportional to the number of threads simultaneously passing through $\gamma_1$ and $\gamma_2$, or in other words, the number of threads connecting $A_1$ and $A_2$, i.e., half of the conditional mutual information between $A_1$ and $A_2$~.

This implies the necessity to design entanglement beyond bipartite. The reason can be seen from a simple analysis. As shown in Fig.~\ref{fig5.1.1}, let's imagine another thread ${\varsigma _{{y_1}{y_2}}}$ connecting points $y_1$ and $y_2$ outside region $A$. ${\varsigma _{{y_1}{y_2}}}$ and ${\varsigma _{{x_1}{x_2}}}$ are superficially overlapping. Asking for the entanglement entropy between ${x_1} \cup {x_2}$ and ${y_1} \cup {y_2}$, we first write down the state of ${x_1} \cup {x_2} \cup {y_1} \cup {y_2}$ as a whole. According to the thread-state rules, if we assume that there is no entanglement between the two threads ${\varsigma _{{y_1}{y_2}}}$ and ${\varsigma _{{x_1}{x_2}}}$, then
\be\label{dir}\left| {{x_1}{x_2}{y_1}{y_2}} \right\rangle  = \frac{1}{{\sqrt 2 }}(\left| {{0_{{x_1}}}{0_{{x_2}}}} \right\rangle  + \left| {{1_{{x_1}}}{1_{{x_2}}}} \right\rangle ) \otimes \frac{1}{{\sqrt 2 }}(\left| {{0_{{y_1}}}{0_{{y_2}}}} \right\rangle  + \left| {{1_{{y_1}}}{1_{{y_2}}}} \right\rangle ).\ee
Obviously, the entanglement entropy between ${x_1} \cup {x_2}$ and ${y_1} \cup {y_2}$ is 0 in (\ref{dir}). The key is that (\ref{dir}) is not symmetric for the four indices $x_1, x_2, y_1, y_2$. If we consider the entanglement entropy between a single index and the other three indices, we get $\log 2$. However, if we consider the entanglement entropy between two specified indices and the other two, different situations arise. For example, considering the entanglement entropy between $x_1\cup y_1$ and $x_2\cup y_2$, we get $2\log 2$, while considering the case between ${x_1} \cup {x_2}$ and ${y_1} \cup {y_2}$ results in 0.

That the thread ${\varsigma _{{x_1}{x_2}}}$ represents the internal entanglement of $A = A_1 \cup A_2$ is equivalent to saying that ${\varsigma _{{x_1}{x_2}}}$ and ${\varsigma _{{y_1}{y_2}}}$ have no entanglement between them. To obtain the correct entanglement entropy $S(A)$ as the sum of the areas of $\gamma_1$ and $\gamma_2$, we should hope ${x_1} \cup {x_2}$ as a whole to provide $2\log 2$ of entanglement with ${y_1} \cup {y_2}$. Therefore, we should modify the state (\ref{dir}) to be completely symmetric for all four indices. As reviewed in Section \ref{sec3}, perfect tensor states can satisfy this requirement. Let us recopy it as follows:
\be\begin{array}{l}
\left| {{x_1}{x_2}{y_1}{y_2}} \right\rangle  = \frac{1}{3}(\left| {{0_{{x_1}}}{0_{{x_2}}}{0_{{y_1}}}{0_{{y_2}}}} \right\rangle  + \left| {{1_{{x_1}}}{1_{{x_2}}}{1_{{y_1}}}{0_{{y_2}}}} \right\rangle  + \left| {{2_{{x_1}}}{2_{{x_2}}}{2_{{y_1}}}{0_{{y_2}}}} \right\rangle \\
\quad \quad \quad \quad \;\,\,\, + \left| {{0_{{x_1}}}{1_{{x_2}}}{2_{{y_1}}}{1_{{y_2}}}} \right\rangle  + \left| {{1_{{x_1}}}{2_{{x_2}}}{0_{{y_1}}}{1_{{y_2}}}} \right\rangle  + \left| {{2_{{x_1}}}{0_{{x_2}}}{1_{{y_1}}}{1_{{y_2}}}} \right\rangle \\
\quad \quad \quad \quad \,\;\,\, + \left| {{0_{{x_1}}}{2_{{x_2}}}{1_{{y_1}}}{2_{{y_2}}}} \right\rangle  + \left| {{1_{{x_1}}}{0_{{x_2}}}{2_{{y_1}}}{2_{{y_2}}}} \right\rangle  + \left| {{2_{{x_1}}}{1_{{x_2}}}{0_{{y_1}}}{2_{{y_2}}}} \right\rangle )
\end{array}.\ee

More thoroughly, to address the issue of characterizing the entanglement entropies of disconnected regions, we can propose a natural picture: in the original thread configuration, although bulk threads may appear to ``intersect’’ with each other, they are actually not coupled with each other. This means that the state of the thread configuration is simply the direct product of the quantum states corresponding to these threads. Now imaging the threads coupling with each other at all intersection points (see Fig.\ref{fig5.1.2}), the original thread configuration becomes a tensor network composed of rank-four tensors. More explicitly, we assumes that these rank-four tensors are recognized as perfect tensors. This is actually a concrete implementation of the idea proposed recently in~\cite{Lin:2023orb}. 

A natural way to argue for this approach is to use recursive thinking: At the beginning, in the original ``direct -product state" picture, we have already successfully characterized the entanglement entropies of all connected subregions. Then we can gradually couple one ``internal thread’’ (such as ${\varsigma _{{x_1}{x_2}}}$) with one ``external thread’’ (such as ${\varsigma _{{y_1}{y_2}}}$) for specified disconnected subregions (such as $A = {A_1} \cup {A_2}$) to further characterize the entanglement entropies of the specified disconnected regions, while without changing other entanglement structures. By adding entanglement into all threads at all ``intersections" in this step-by-step way, it can be expected that the entanglement entropies of all disconnected regions can be consistently characterized. We will provide a more rigorous proof in the next section.

The issue of characterizing the entanglement entropies of disconnected regions also suggests that kinematic space cannot be simply related to a simple direct product of the quantum states of each spatial point, even though in kinematic space, the entropy corresponding to diamond-shaped region is notably proportional to its volume. This is because now the threads one-to-one mapped with the spatial points in kinematic space should be understood as being entangled with each other. Therefore, points in kinematic space are also highly entangled with each other.


\subsection{ Proof}\label{sec52}
\begin{figure}
    \centering
    \includegraphics[scale=0.7]{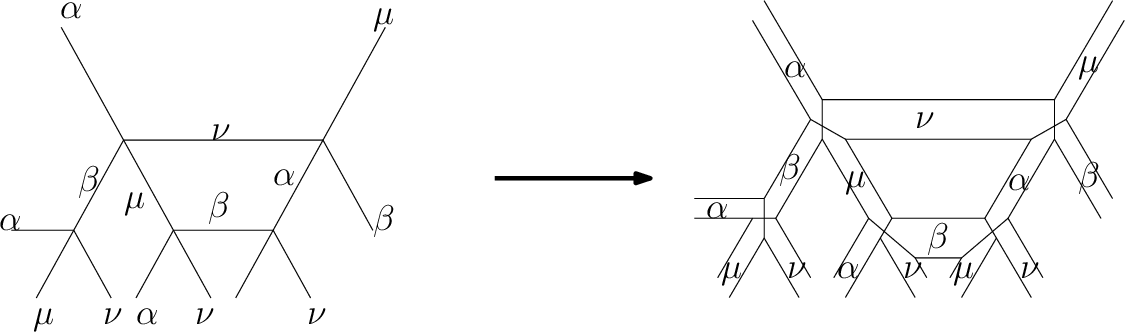}
    \caption{The ``Four Color Theorem" ensures that index types $\alpha, \beta, \mu, \text{or} \nu$ can be unambiguously assigned to each leg in the network.}
    \label{fig5.2.1}
\end{figure}

Firstly, we point out that assigning a rank-four perfect tensor state to each ``four-legged subdiagram" in the ``entangled thread configuration" and eventually unambiguously gluing these small tensors together is feasible. We agree to label each rank-four perfect tensor as ${T_{\alpha \beta \mu \nu}}$, where the order of the four qutrits $\alpha, \beta, \mu, \nu$ is important. In other words, once we specify the four legs of a four-legged subdiagram as being respectively associated with $\alpha, \beta, \mu, \nu$, the entangled state of the four qutrits as a whole is defined according to the pattern specified in Eq.~(\ref{ksi}). Now, contraction implies index summation, so to avoid ambiguity, we stipulate that when two adjacent four-legged subdiagrams are glued, legs with the same label are always connected, as shown in Fig.~\ref{fig5.2.1}. Thus, when dealing with an entire tensor network diagram glued together by multiple four-legged diagrams, the unambiguous requirements for tensor contraction are equivalent to the need to assign index types $\alpha, \beta, \mu, \text{or} \nu$ to each leg in the tensor network, ensuring that legs sharing a vertex cannot be assigned the same index types. Fortunately, this is explicitly achievable. Considering each index type as a color and each leg in the network as an area with a certain width, what we require is simply to color a ``map" with four different colors in such a way that any two adjacent regions must be painted in different colors. The well-known ``Four Color Theorem" in mathematics has already told us that this intuition is precisely correct.

\subsubsection{Calculation of Entanglement Entropy for A Connected Region}
\begin{figure}
    \centering
    \includegraphics[scale=1.1]{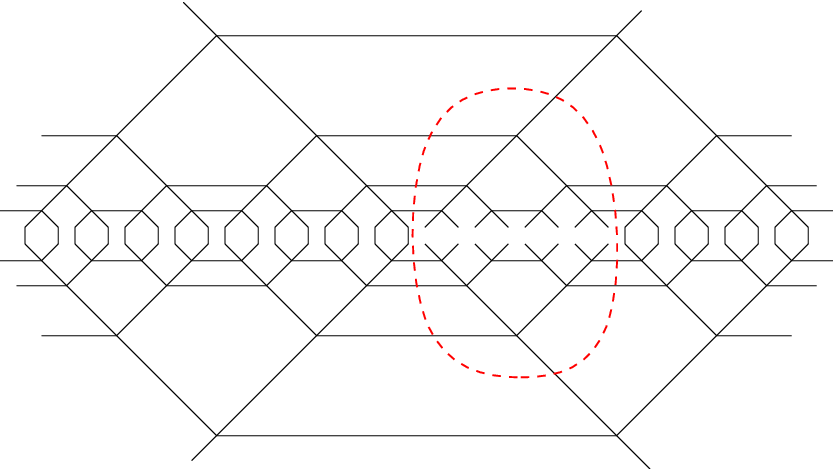}
    \caption{The tensor network diagrammatic representation of the expression (\ref{rhoa}), where taking the trace over $\bar A$ means gluing together the indices inside $\bar A$.}
    \label{fig5.2.2}
\end{figure}

\begin{figure}
    \centering
    \includegraphics[scale=0.5]{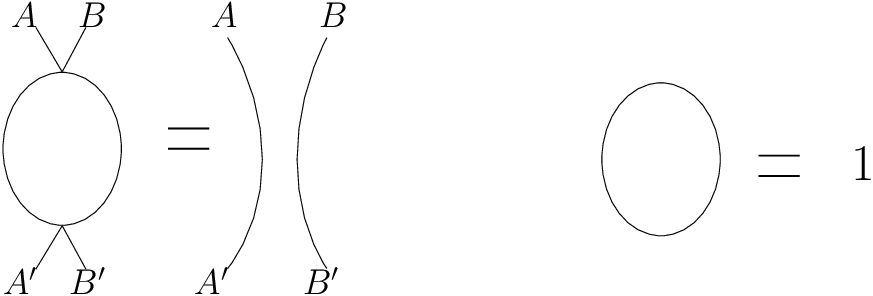}
    \caption{The diagram representation of rules (\ref{del}) and (\ref{cir}).}
    \label{fig5.2.3}
\end{figure}

We first prove that in this ``entangled thread configuration" with assigned perfect tensor states, the calculation of the entanglement entropy $S(A)$ for a single connected region $A$ satisfies the holographic RT prescription. To do this, we first calculate the reduced density matrix
\be\label{rhoa}{\rho_A} = \text{tr}_{\bar A}(\left| \Psi \right\rangle \left\langle \Psi \right|),\ee
where $\left| \Psi \right\rangle$ is now the entire tensor network state formed by gluing all perfect tensor states. Fig.~\ref{fig5.2.2} illustrates the tensor network diagrammatic representation of the expression (\ref{rhoa}), where taking the trace over $\bar A$ means gluing together the indices inside $\bar A$. Now, we will utilize the following interesting properties of perfect tensors (depicted in Fig.~\ref{fig5.2.3}),
\be\label{del}\sum\limits_{\alpha \beta } {{T_{\alpha \beta \mu \nu }}{{({T_{\alpha \beta \mu '\nu '}})}^*}}  = \frac{1}{9}{\delta _{\mu \mu '}}{\delta _{\nu \nu '}},\ee
and
\be\label{cir}\sum\limits_\mu  {\frac{1}{3}{\delta _{\mu \mu '}}}  = 1,\ee
where each $1/3 \delta$ represents a tensor $\left[ {\begin{array}{*{20}{c}} {\frac{1}{3}}&0&0\\ 0&{\frac{1}{3}}&0\\ 0&0&{\frac{1}{3}} \end{array}} \right]$, characterizing the density matrix of a single qutrit as $\rho = \sum\limits_{i = 0,1,2} {\frac{1}{3}\left| i \right\rangle \left\langle i \right|}$, and thus conveniently represented by a single thread. A circle then represents the trace over the tensor $1/3 \delta$, yielding 1.

\begin{figure}
     \centering
       \begin{subfigure}[b]{0.4\textwidth}
         \centering
         \includegraphics[width=\textwidth]{5.2.2.eps}
         \caption{}
         \label{fig5.2.0}
     \end{subfigure}
     \hfill
     \begin{subfigure}[b]{0.4\textwidth}
         \centering
         \includegraphics[width=\textwidth]{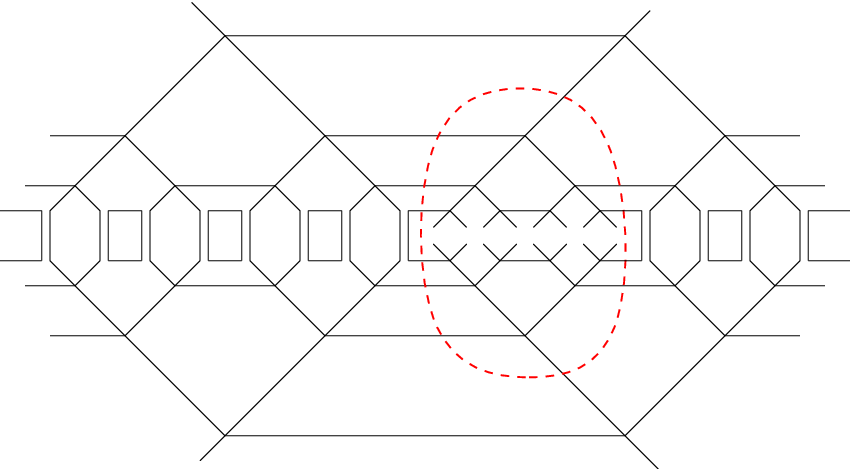}
         \caption{}
         \label{fig5.2.4}
     \end{subfigure}
     \hfill
     \begin{subfigure}[b]{0.4\textwidth}
         \centering
         \includegraphics[width=\textwidth]{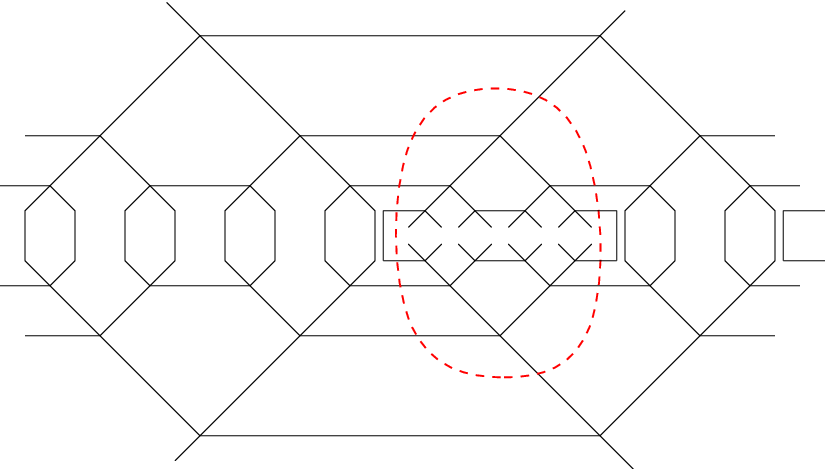}
         \caption{}
         \label{fig5.2.5}
     \end{subfigure}
     \hfill
     \begin{subfigure}[b]{0.4\textwidth}
         \centering
         \includegraphics[width=\textwidth]{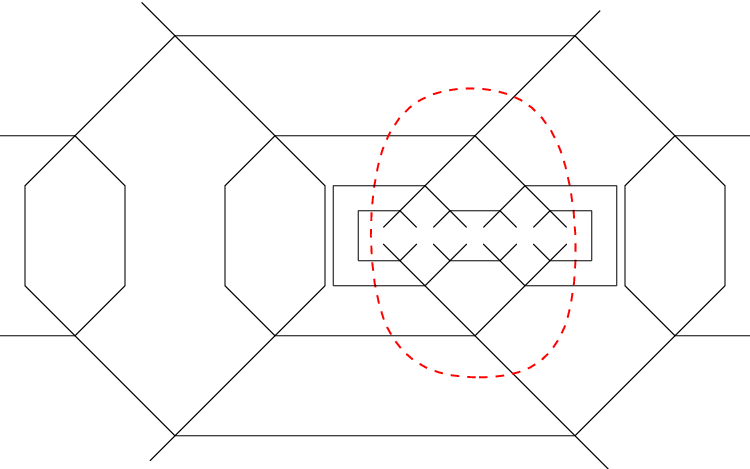}
         \caption{}
         \label{fig5.2.7}
     \end{subfigure}
           \hfill
     \begin{subfigure}[b]{0.4\textwidth}
         \centering
         \includegraphics[width=\textwidth]{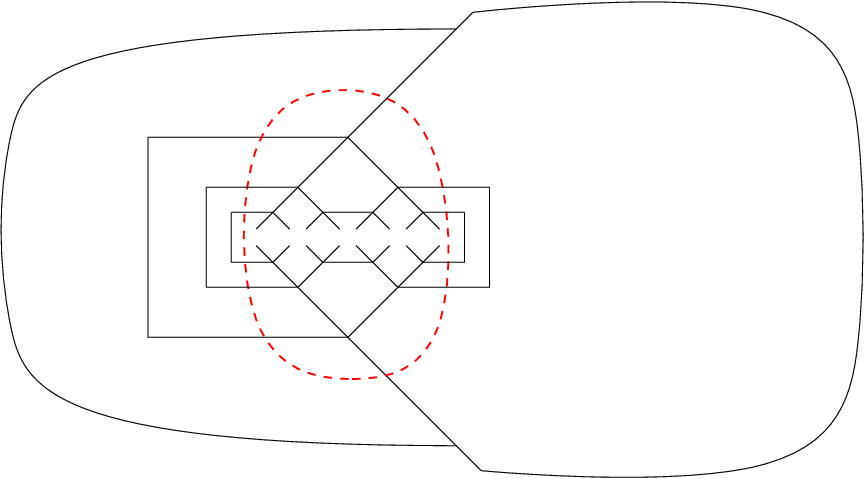}
         \caption{}
         \label{fig5.2.8}
     \end{subfigure}
        \caption{(a) The original diagram representing (\ref{rhoa}); (b) The first step, in which (\ref{del}) has been used; (c) The second step, in which (\ref{cir}) has been used; (d) Iteratively using (\ref{del}) and (\ref{cir}) due to the symmetry; (e) The net result.}
        \label{fig5.2.4-8}
\end{figure}

Applying these two simple rules iteratively, we can simplify the expression (\ref{rhoa}) step by step, and all operations will be performed graphically, see Fig.~\ref{fig5.2.4-8}. In the first step, we use Eq.~(\ref{del}), and in the second step, we use Eq.~(\ref{cir}). By combining these two steps, we find that this process suggests that short-range entanglement from previous layers has not entered into the calculation at larger scales. Moreover, after this process, a symmetric pattern suitable for iteration emerges. We can use (\ref{del}) and (\ref{cir}) again to further remove short-range entanglement. After repeated iterations, the net result is obtained in Fig.~\ref{fig5.2.8}. Observing this result, to calculate the entanglement entropy of $A$, we actually only need to consider a subnetwork, which can be precisely interpreted as the entanglement wedge $W(A)$ of $A$. Denoting the lattice sites included in $A$ as ${x_a}$ and the sites included in the minimal cut $\gamma_A$ as ${s_\gamma}$, this entanglement wedge subnetwork can be seen as characterizing a pure state of the whole set $\{ {x_a}\}  \cup \{ {s_\gamma }\}$, and calculating the von Neumann entropy of $A$ is also equivalent to calculating the entanglement entropy between $\{ {x_a}\}$ and $\{ {s_\gamma }\}$.
\begin{figure}
     \centering
     \begin{subfigure}[b]{0.4\textwidth}
         \centering
         \includegraphics[width=\textwidth]{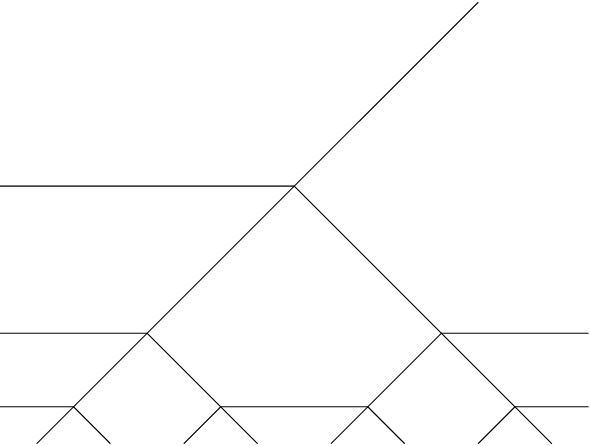}
         \caption{}
         \label{fig5.2.9}
     \end{subfigure}
     \hfill
     \begin{subfigure}[b]{0.3\textwidth}
         \centering
         \includegraphics[width=\textwidth]{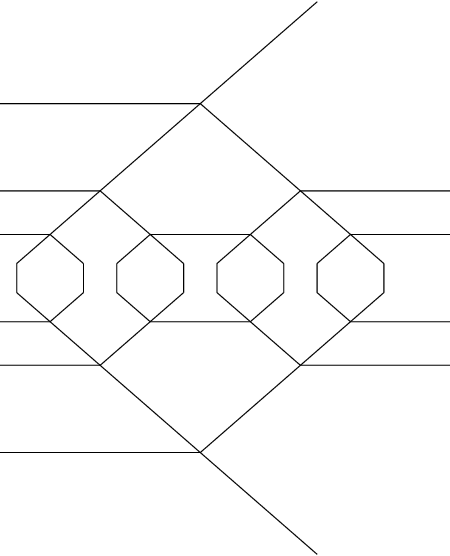}
         \caption{}
         \label{fig5.2.10}
     \end{subfigure}
     \hfill
     \begin{subfigure}[b]{1.0\textwidth}
         \centering
         \includegraphics[width=\textwidth]{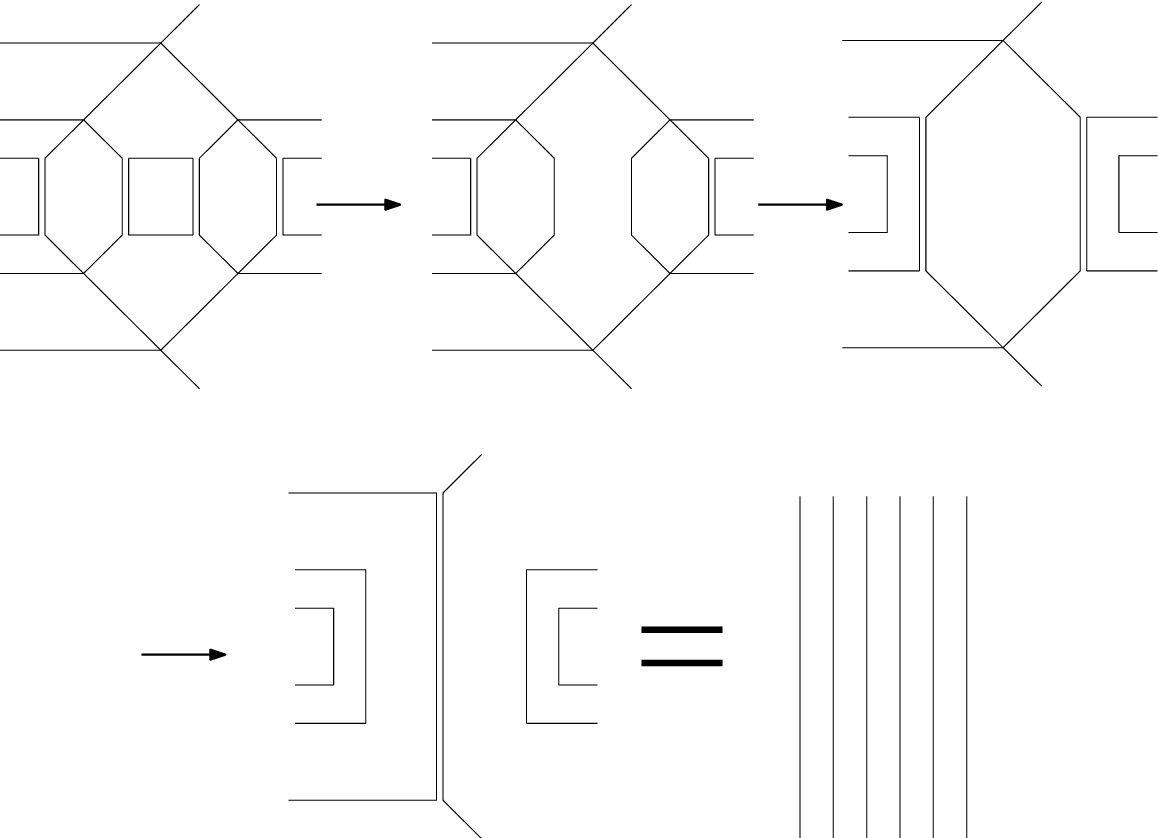}
         \caption{}
         \label{fig5.2.11}
     \end{subfigure}
        \caption{(a) Fig.~\ref{fig5.2.8} tell us that to calculate $S(A)$ we only need to consider a subnetwork. (b) Equivalently calculate the entanglement entropy between $\{ {x_a}\}$ and $\{ {s_\gamma }\}$.  (c) Iteratively applying (\ref{del}) and (\ref{cir}) finally leads to a simple result.}
        \label{}
\end{figure}

To calculate this entanglement entropy more conveniently, as shown in Fig.~\ref{fig5.2.10}, we next choose to contract $\{ {x_a}\}$ rather than $\{ {s_\gamma }\}$. In other words, we calculate the reduced density matrix corresponding to the entanglement entropy
\be{\rho _{{\rm{ent}}}} = \text{tr}_{\{ {x_a}\}}(\left| {{\psi _A}} \right\rangle \left\langle {{\psi _A}} \right|),\ee
where $\left| {{\psi _A}} \right\rangle$ represents the state corresponding to the entanglement wedge subnetwork of $A$.

As shown in Fig.~\ref{fig5.2.11}, we then iteratively apply the expressions (\ref{del}) and (\ref{cir}) again, and finally, we find that the graphical representation of this density matrix is simply a cluster of threads representing the direct product of Bell pairs! Moreover, the number of threads is exactly equal to the number of threads crossed by the minimal cut! Thus, we verify the first expectation that coupling the thread configuration according to the entanglement pattern of perfect tensors does not affect the expression of the entanglement entropy for connected subregions. Note that similarly, in the process of graphical calculation, it can be clearly seen that short-range entanglement within $A$ does not contribute to the entanglement entropy of $A$.


\subsubsection{Calculation of Entanglement Entropies for Disonnected Regions}

Next, we will clarify the proof that resolves the calculation problem of entanglement entropy for disconnected regions proposed in Section \ref{sec51}. The method is similar, employing the expressions (\ref{del}) and (\ref{cir}), with some subtle details that we present in this section.

\begin{figure}
     \centering
     \begin{subfigure}[b]{0.7\textwidth}
         \centering
         \includegraphics[width=\textwidth]{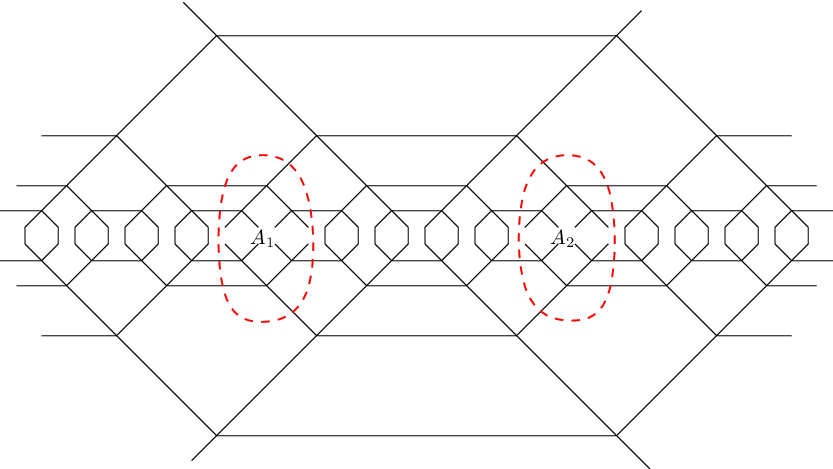}
         \caption{}
         \label{fig5.2.12}
     \end{subfigure}
     \hfill
     \begin{subfigure}[b]{0.4\textwidth}
         \centering
         \includegraphics[width=\textwidth]{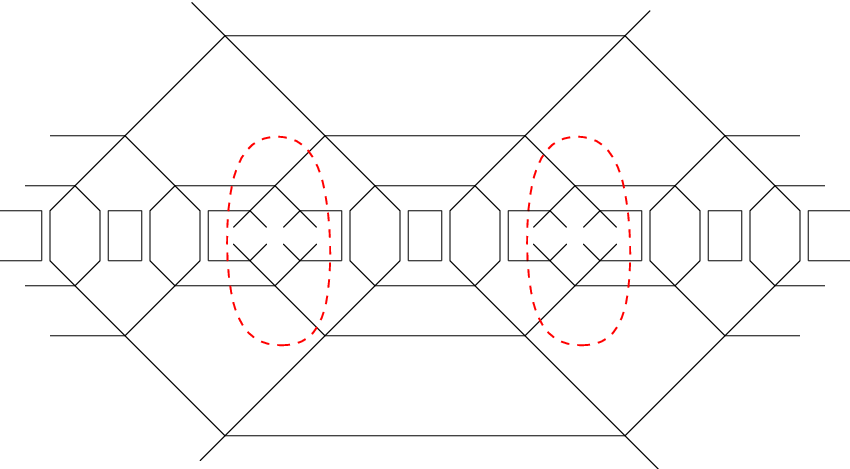}
         \caption{}
         \label{fig5.2.13}
     \end{subfigure}
     \hfill
     \begin{subfigure}[b]{0.4\textwidth}
         \centering
         \includegraphics[width=\textwidth]{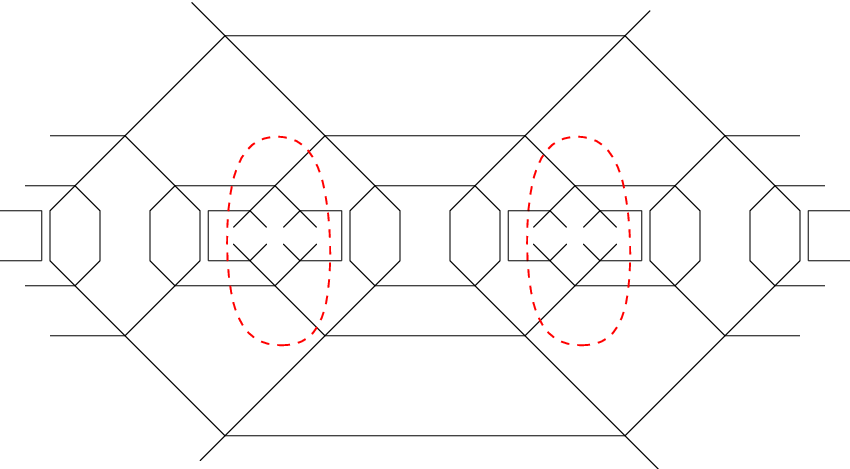}
         \caption{}
         \label{fig5.2.14}
     \end{subfigure}
      \hfill
     \begin{subfigure}[b]{0.8\textwidth}
         \centering
         \includegraphics[width=\textwidth]{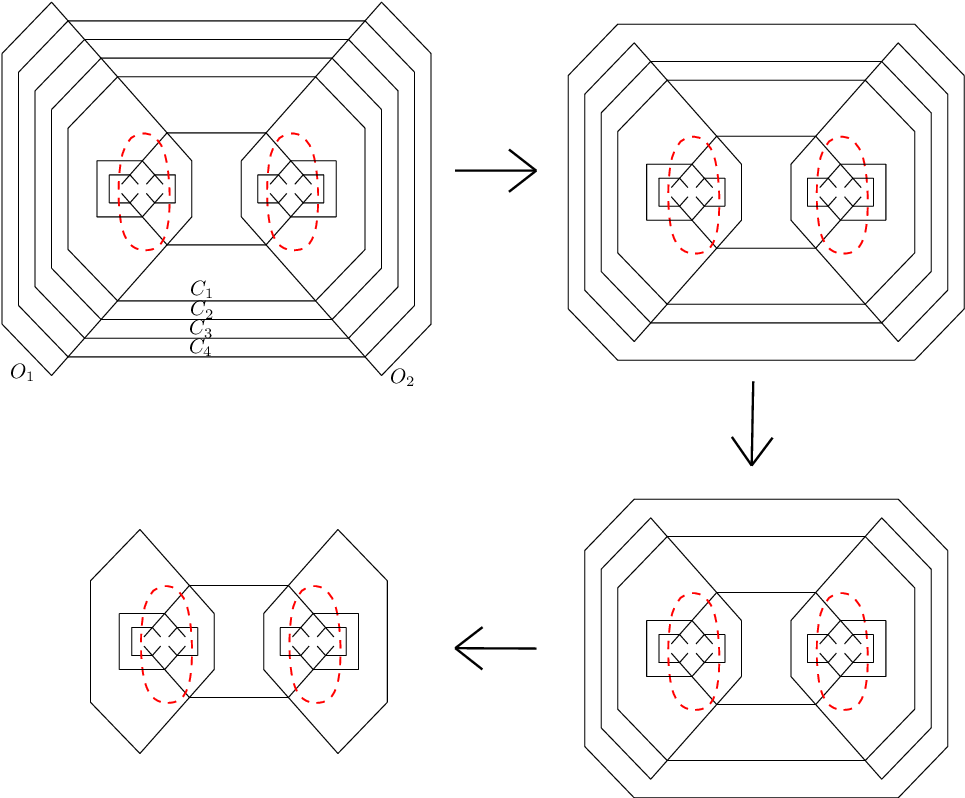}
         \caption{}
         \label{fig5.2.15}
     \end{subfigure}
        \caption{Calculation of entanglement entropy for a disonnected region.}
        \label{fig5.2.13-14}
\end{figure}

\begin{figure}
    \centering
    \includegraphics[scale=0.8]{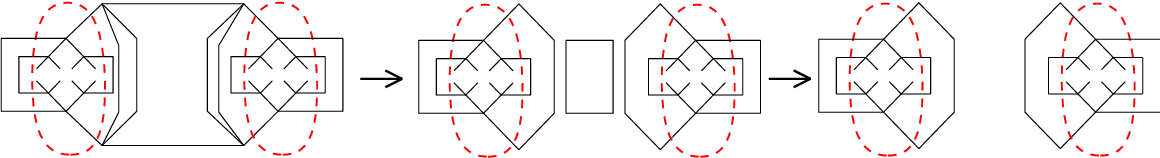}
    \caption{Calculation of entanglement entropy for a disonnected region.}
    \label{fig5.2.16}
\end{figure}

Consider the configuration as shown in Fig.~\ref{fig5.2.12}, where the entanglement entropy of $A = A_1 \cup A_2$ is given by the sum of the areas of the RT surfaces $\gamma_1$ for $A_1$ and $\gamma_2$ for $A_2$. The steps for calculating the entanglement entropy are similar. Again, we first calculate the reduced density matrix $\rho_A$ by gluing together the indices inside $\bar A$. Then, we iteratively apply the two simple rules (\ref{del}) and (\ref{cir}) to simplify the expression (\ref{rhoa}). The first and second steps are illustrated in Fig.~\ref{fig5.2.13-14}. Similarly, symmetry allows an iterative scheme, and we can proceed similarly to Fig.~\ref{fig5.2.11}. It can be verified that by iterating continuously, the final pattern in Fig.~\ref{fig5.2.15} is obtained.

Note that in the current context, although, during the iteration process, entanglement at each scale has been removed by (\ref{del}) and (\ref{cir}), some long-range entanglement crossing the disconnected regions $A_1$ and $A_2$ has not been removed. This is manifested in the pattern as circles around $A_1$ and $A_2$. Let's denote these circles from inner to outer as $C_1, C_2, \ldots, C_n$, where these circles intersect only at circles $O_1$ surrounding $A_1$ or circles $O_2$ surrounding $A_2$. Note that the more outer circles have already been removed by the previous standard procedure through (\ref{del}) and (\ref{cir}).

However, we can further eliminate circles $C_1, C_2, \ldots, C_n$ using (\ref{del}) and (\ref{cir}). As shown in Fig.~\ref{fig5.2.16}, we first apply (\ref{del}) at the intersection of the outermost circle $C_n$ with either $O_1$ or $O_2$, and the net result is the decoupling of the circle $C_n$ from $O_1$ and $O_2$, i.e., they no longer intersect. Therefore, by (\ref{cir}), $C_n$ no longer contributes to the entropy calculation. Obviously, this process can be repeated, successively removing $C_{n-1}, \ldots, C_1$, until reaching the final configuration in Fig.~\ref{fig5.2.15}.

To handle the final configuration in Fig.~\ref{fig5.2.15}, note that it is topologically equivalent to the configuration (a) in Fig.~\ref{fig5.2.16}. Thus, as shown in Fig. \ref{fig5.2.16}, we can first apply (\ref{del}) and then (\ref{cir}) to reach the configuration (c) in Fig.~\ref{fig5.2.16}. However, configuration(c) is nothing else but the sum of the entanglement entropy for regions $A_1$ and $A_2$, and the entanglement entropy for $A_1$ and $A_2$ is precisely given by the number of cuts through their respective entanglement wedges. We can replicate the calculation in Fig.~\ref{fig5.2.11} and obtain the following result:

\be S({A_1} \cup {A_2}) = k\ln 3(\# \text{cut}_{\gamma ({A_1})} + \# \text{cut}_{\gamma ({A_2})}).\ee

Thus, we obtain

\be\label{plus}S({A_1} \cup {A_2}) = S({A_1}) + S({A_2}),\ee

where $\# \text{cut}_{\gamma ({A_1})}$ represents the number of cut legs through the minimal surface $\gamma ({A_1})$. $k$ is a coefficient chosen as needed.

Note that, according to this, the paradox proposed in Section \ref{sec31} has received an insightful resolution. In our case, there indeed exists a thread (the green thread shown in Fig.~\ref{fig5.2.12}) that characterizes the conditional mutual information (CMI) between regions $A_1$ and $A_2$. If we simply interpret this CMI as a kind of bipartite entanglement, we cannot consistently characterize (\ref{plus}). But now we consider this thread as being in an entangled state with other threads, and we have just proven that this entanglement scheme allows the characterization of entanglement entropy for disconnected regions.

\begin{figure}
     \centering
     \begin{subfigure}[b]{0.7\textwidth}
         \centering
         \includegraphics[width=\textwidth]{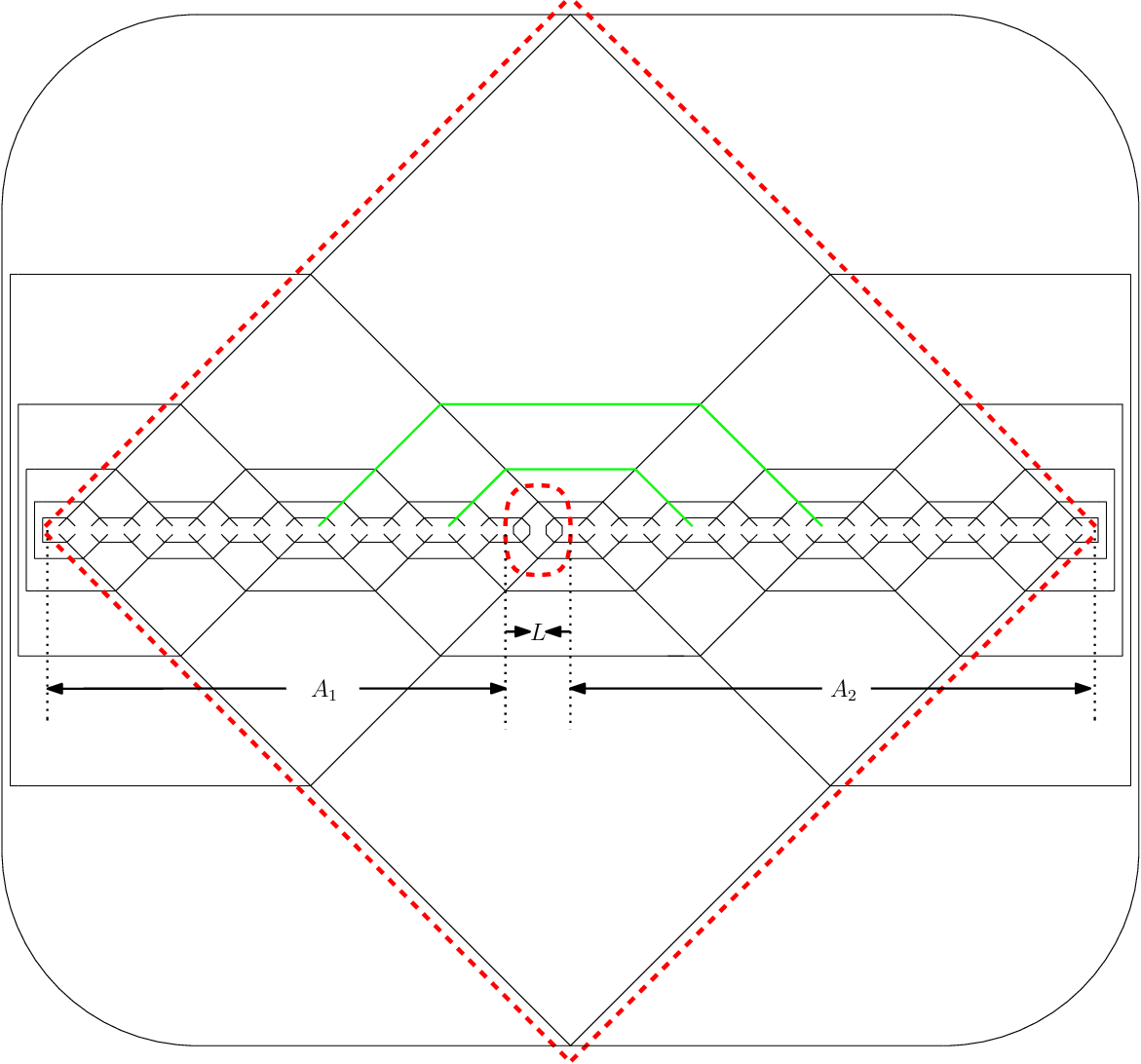}
         \caption{}
         \label{fig5.2.17}
     \end{subfigure}
     \hfill
     \begin{subfigure}[b]{0.7\textwidth}
         \centering
         \includegraphics[width=\textwidth]{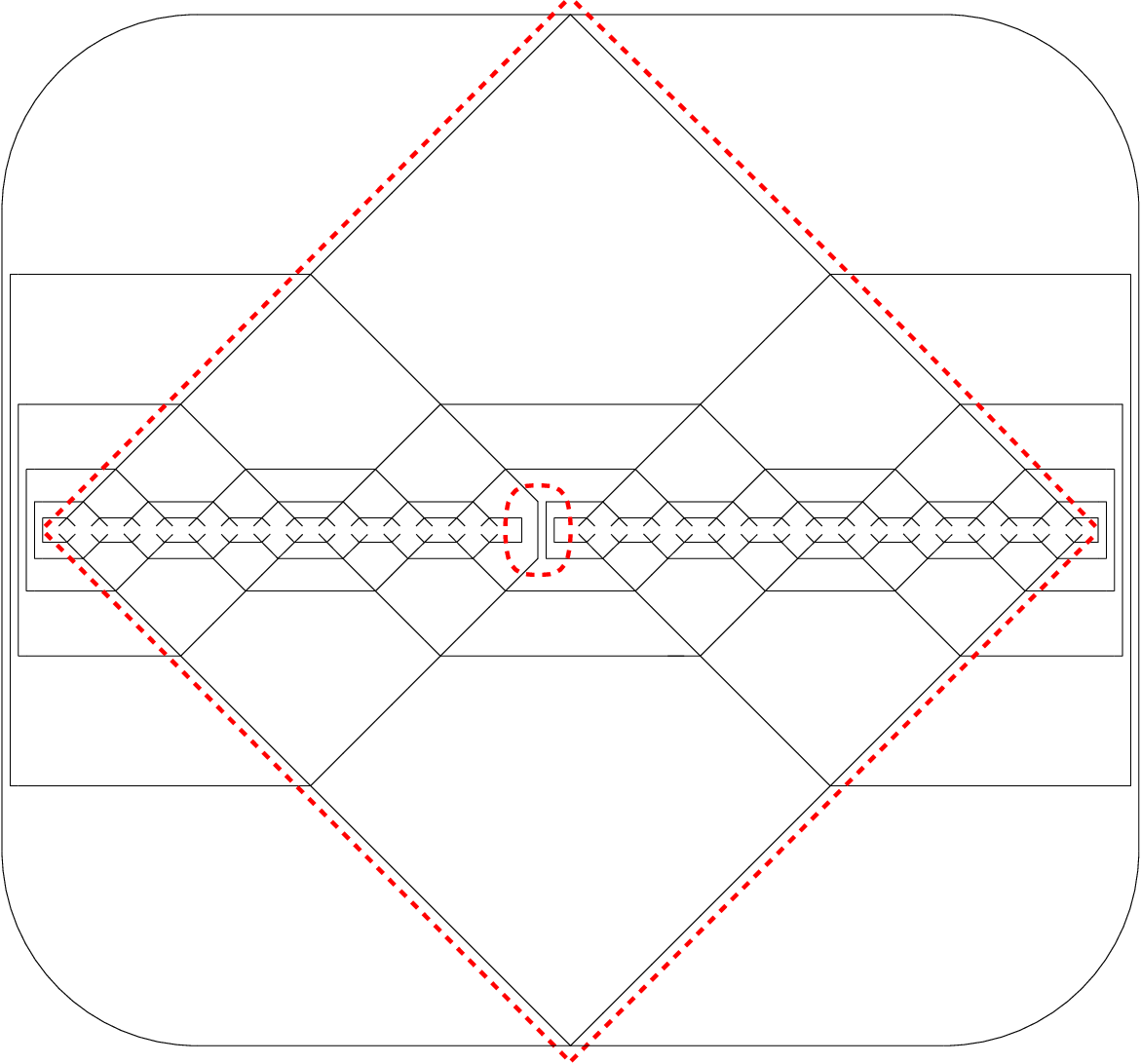}
         \caption{}
         \label{fig5.2.18}
     \end{subfigure}     
        \caption{(a) The reduced density matrix for another case of disconnected region’s entanglement entropy. (b) Firstly reduce the thread configuration inside $\gamma (L)$.}
        \label{}
\end{figure}

Now, for completeness, we should also prove that this perfect tensor-type thread configuration can give another facet of entropy for disconnected subregions. As is well known, for a disconnected subregion $A = A_1 \cup A_2$, the holographic RT surface can be either $\gamma ({A_1}) \cup \gamma ({A_2})$ or $\gamma ({A_1} \cup L \cup {A_2}) \cup \gamma (L)$, where $L$ is the intermediate region between $A_1$ and $A_2$, as long as the latter gives a smaller area. Fig.~\ref{fig5.2.17} shows an example of this situation, noting that there are still threads connecting $A_1$ and $A_2$ that characterize the conditional mutual information relative to $L$. It now needs to be proved that in this case, counting the number of cut legs still correctly gives the entanglement entropy, i.e., prove:

\be S({A_1} \cup {A_2}) = S({A_1} \cup L \cup {A_2}) + S(L),\ee

or

\be S({A_1} \cup {A_2}) = k\ln 3(\# \text{cut}_{\gamma ({A_1} \cup L \cup {A_2})} + \# \text{cut}_{\gamma (L)}).\ee

\begin{figure}
     \centering
     \begin{subfigure}[b]{0.8\textwidth}
         \centering
         \includegraphics[width=\textwidth]{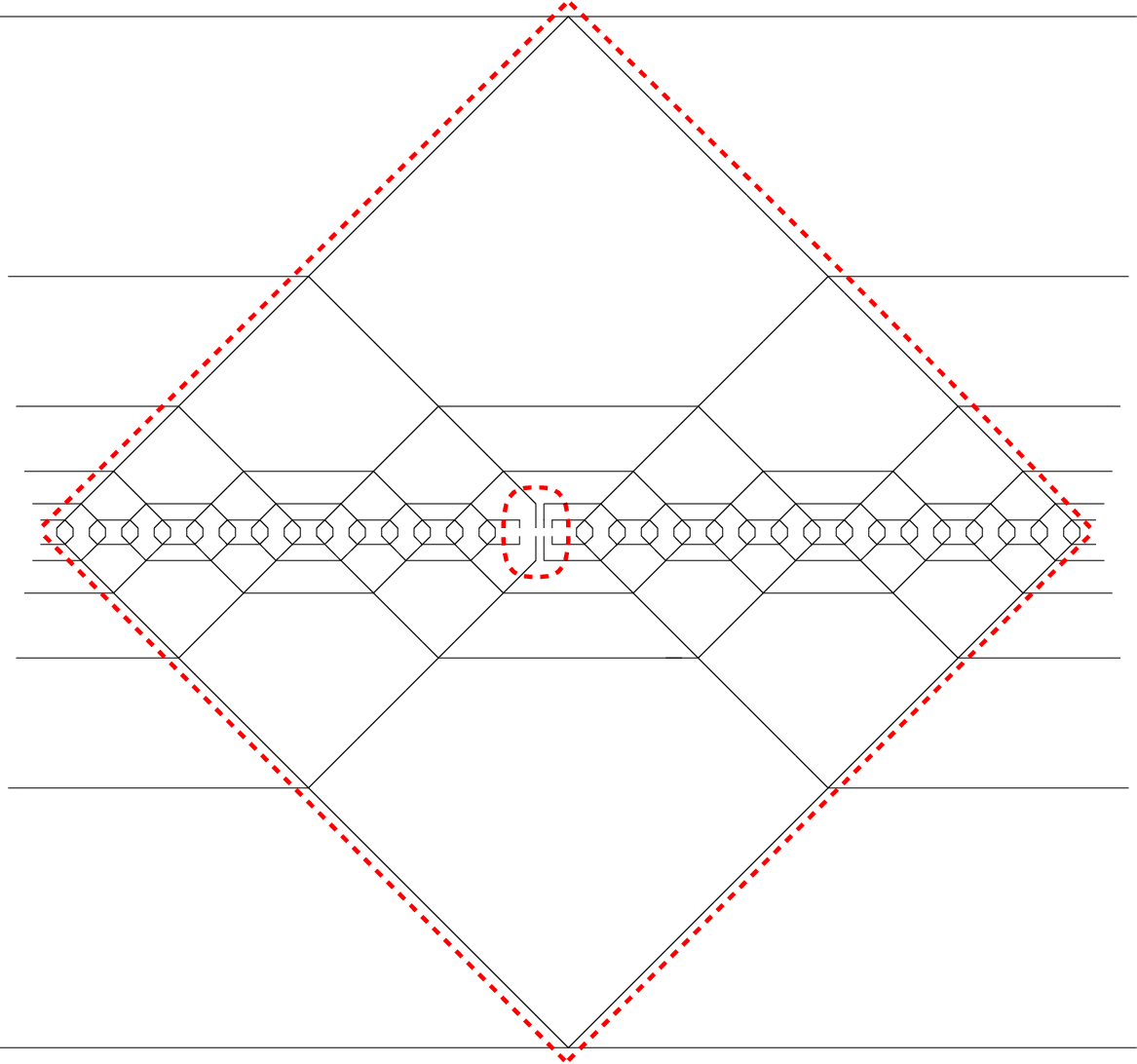}
         \caption{}
         \label{fig5.2.19}
     \end{subfigure}
     \hfill
     \begin{subfigure}[b]{0.5\textwidth}
         \centering
         \includegraphics[width=\textwidth]{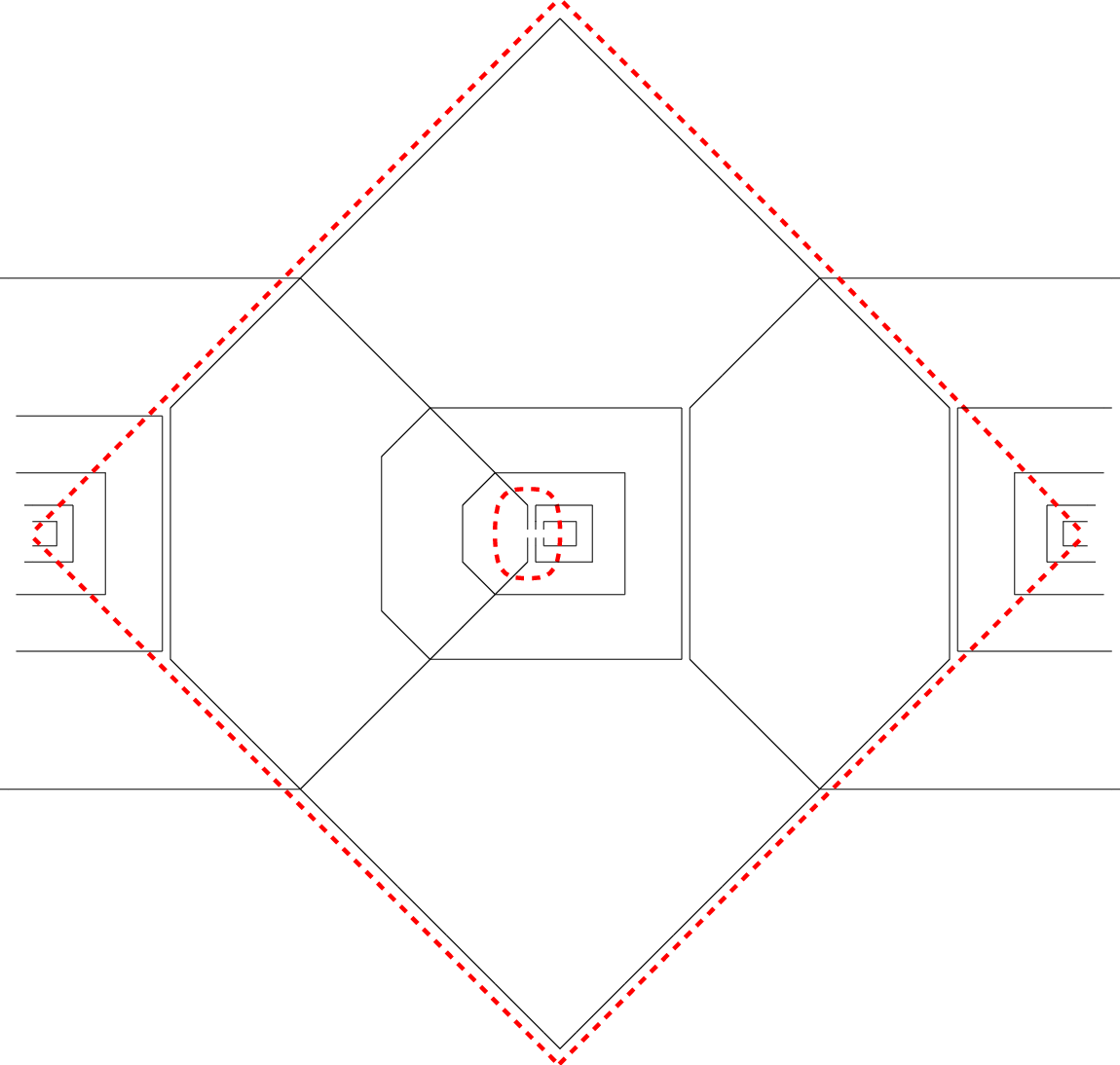}
         \caption{}
         \label{fig5.2.20}
     \end{subfigure}     
     \hfill
     \begin{subfigure}[b]{0.45\textwidth}
         \centering
         \includegraphics[width=\textwidth]{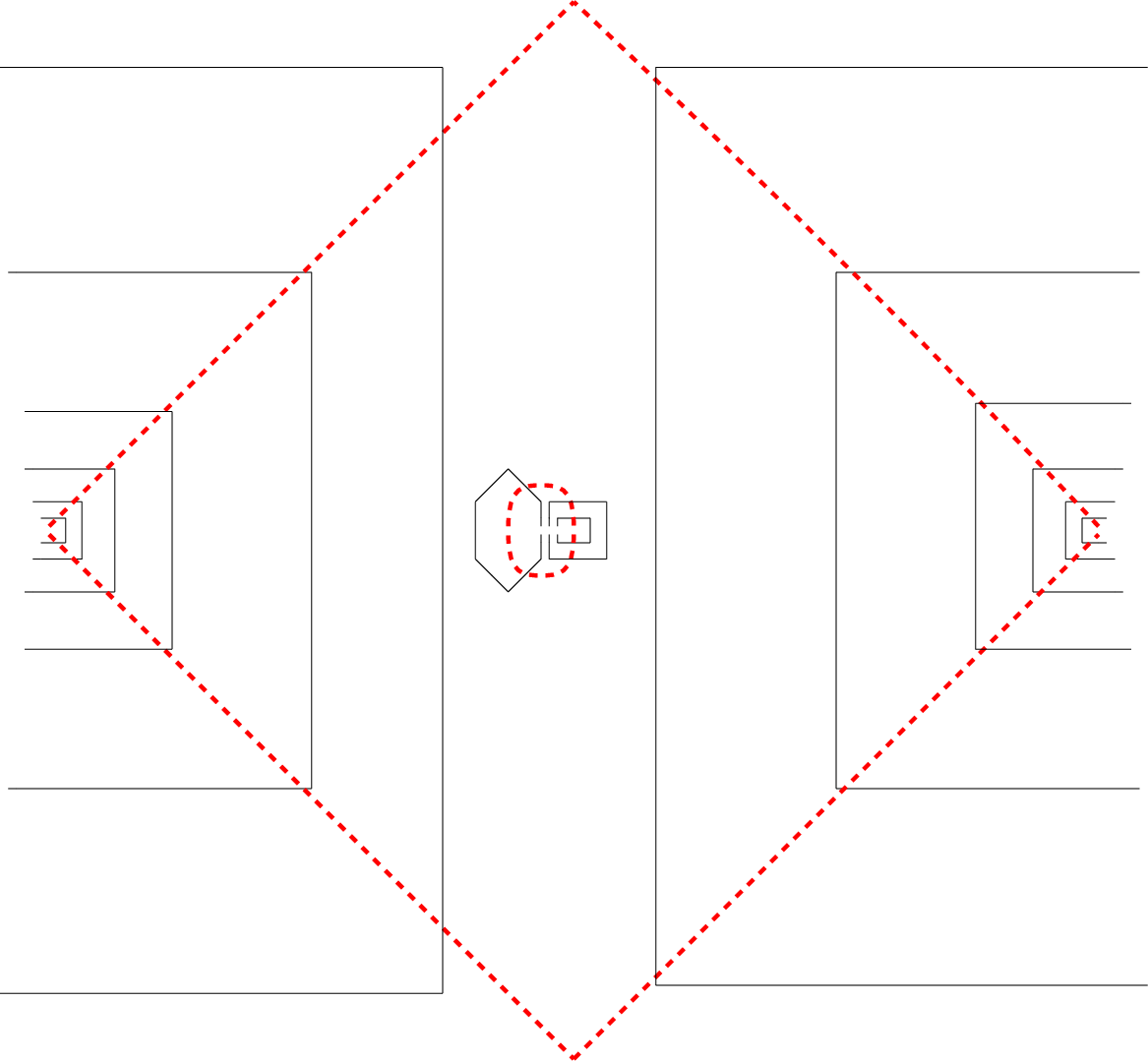}
         \caption{}
         \label{fig5.2.22}
     \end{subfigure}  
        \caption{The calculation details. }
        \label{}
\end{figure}

Based on the previous calculation experience, we can now easily go from the original configuration of calculating the reduced density matrix $\rho_A$ directly to the pattern in Fig.~\ref{fig5.2.17}. This is because we can continuously use (\ref{del}) and (\ref{cir}) to reduce the threads and circles outside the surface $\gamma ({A_1} \cup L \cup {A_2})$ to the configuration in Fig.~\ref{fig5.2.17}. On the other hand, from the calculation in Fig.~\ref{fig5.2.3}, we can also directly reduce the thread configuration inside the surface $\gamma (L)$ to the configuration in Fig.~\ref{fig5.2.18}.

Following a similar logic to step~\ref{fig5.2.10}, now Fig.~\ref{fig5.2.18} is equivalent to calculating Fig.~\ref{fig5.2.19}. Thus, we arrive at a pattern that is very easy to directly apply (\ref{del}) and (\ref{cir}) for reduction, and after several steps of operation, we get the pattern in Fig.~\ref{fig5.2.20}. Making a deformation that preserves the topology, we obtain the pattern in Fig.~\ref{}. With this, we can continue to use (\ref{del}) and (\ref{cir}), thus finally obtaining Fig.~\ref{fig5.2.22}. Fig.~\ref{fig5.2.22} consists of two parts, where the smaller part, by a calculation similar to Fig.~\ref{fig5.2.8}, will contribute to $S(L)$, and the other part precisely calculates the entanglement entropy of the reduced density matrix represented by $n$ threads, where $n$ is precisely proportional to $\gamma ({A_1} \cup L \cup {A_2})$. Thus, we have completed the proof.



\section{ Characterization of Entanglement Wedge Cross-Sections}


\subsection{Review of Entanglement of Purification}

In the holographic duality, there exist many quantum information theory quantities that characterize the correlations between subsystems ${A_1}$ and ${A_2}$ in a mixed-state bipartite system $A = {A_1} \cup {A_2}$. Here, we choose to use the entanglement of purification ${E_P}({A_1}:{A_2})$ (EoP)~\cite{Nguyen:2017yqw,Takayanagi:2017knl} to characterize this correlation. The method is as follows: imagine introducing two auxiliary systems, denoted as ${A'_1}$ and ${A'_2}$, so that ${A_1} \cup {A_2} \cup {A'_1} \cup {A'_2}$ as a whole is in a pure state $\psi ({A_1}{A_2}{A'_1}{A'_2})$. In this way, one can legitimately define the entanglement entropy between ${A_1}{A'_1}$ and ${A_2}{A'_2}$. As there are infinitely many purification schemes, the entanglement of purification takes the minimum entanglement entropy $S({A_1}{A'_1})$ between ${A_1}{A'_1}$ and ${A_2}{A'_2}$ over all possible schemes as the correlation between ${A_1}$ and ${A_2}$, i.e.,
\be{E_P}({A_1}:{A_2}) = \mathop {\min }\limits_{{{\left| \psi  \right\rangle }_{{A_1}{{A'}_1}{A_2}{{A'}_2}}}} S({A_1}{A'_1}).\ee

Similar to the RT formula, it has been proposed in the literature~\cite{Nguyen:2017yqw,Takayanagi:2017knl} that the entanglement of purification can be calculated from the area of the dual EWCS surface, i.e.,
\be{E_P}({A_1}:{A_2}) = \frac{{{\rm{Area}}\left( {{\sigma _{{A_1}:{A_2}}}} \right)}}{{4{G_N}}}.\ee

Detailed discussions on holographic entanglement of purification are found in the literature on surface-state duality (see~\cite{Bao:2018fso,Bao:2018zab,Du:2019emy}). Since what we want to measure is the (minimal possible) entanglement entropy between ${A_1}{A'_1}$ and ${A_2}{A'_2}$, the optimal scheme should first satisfy that the auxiliary system ${A'_1} \cup \;{A'_2}$ itself has no internal entanglement, allowing its Hilbert space to be used most economically, thus providing a purification with the smallest possible Hilbert space dimension. Intuitively, the dimension of the Hilbert space of ${A'_1} \cup \;{A'_2}$ with no internal entanglement is precisely used entirely to characterize the entanglement between ${A_1}$ and ${A_2}$, thereby capturing the intrinsic correlation between ${A_1}$ and ${A_2}$.

\subsection{Necessity of Perfect Entanglement for Mixed-State Intrinsic Correlations}

\begin{figure}
    \centering
    \includegraphics[scale=0.82]{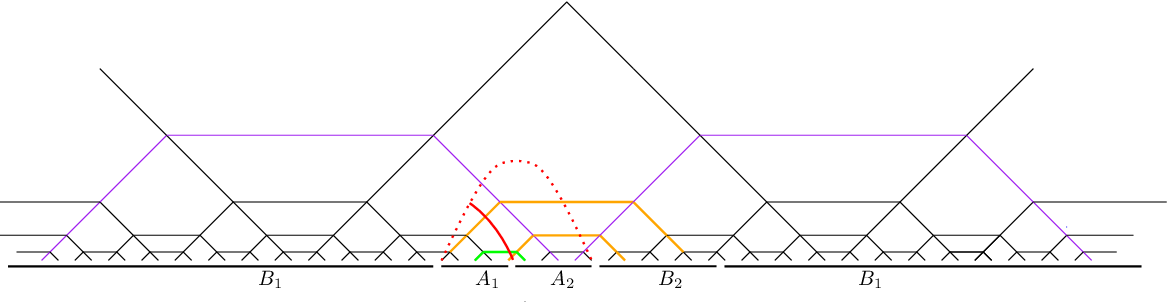}
    \caption{Here the entanglement wedge cross-section is the minimal cut indicated by the red solid line. We have arranged for the number of yellow threads to be equal to the number of purple threads, which means ${N_{{A_1} \leftrightarrow {B_2}}} = {N_{{A_2} \leftrightarrow {B_1}}}$.}
    \label{fig6.2.1}
\end{figure}

Let's choose $A_1$ and $A_2$ as shown in the Fig~\ref{fig6.2.1}, and we have marked the types of various threads with colors in the figure. Before ``tying up" all the threads (corresponding to the original direct-product coarse-grained state), we can see that the green threads represent the connections between $A_1$ and $A_2$, the yellow threads represent the connections from $A_1$ to the region we define as $B_2$, and the purple threads represent the connections from $A_2$ to the region we define as $B_1$. Note that we have arranged for the number of yellow threads to be equal to the number of purple threads. In other words, we have agreed on $B_1$ and $B_2$ in such a way that
\be{N_{{A_1} \leftrightarrow {B_2}}} = {N_{{A_2} \leftrightarrow {B_1}}}.\ee
Thus, we obtain a configuration consistent with Section \ref{sec32}. We can also precisely draw the position of the entanglement wedge cross-section in the current context. It is the minimal cut (indicated by red solid line in the figure) that starts from the boundary point of $A_1$ and $A_2$, and divides the entanglement wedge $W({A_1} \cup {A_2})$ into two halves. In this setup, we can reproduce the ``experimental results" (\ref{sig}):
Given
\be{N_{{A_1} \leftrightarrow {A_2}}} = 1,\ee
\be{N_{{A_1} \leftrightarrow {B_2}}} = {N_{{A_2} \leftrightarrow {B_1}}} = 2,\ee
\be\label{ans}\frac{{{\rm{Area}}\left( {{\sigma _{{A_1}:{A_2}}}} \right)}}{{4{G_N}}} = \# \min {\rm{imal}}\;{\rm{cut}} = 3,\ee
we have
\be\label{equ2}\frac{{Area\left( {{\sigma _{{A_1}:{A_2}}}} \right)}}{{4{G_N}}} = {N_{{A_1} \leftrightarrow {B_2}}} + {N_{{A_1} \leftrightarrow {A_2}}} = {N_{{A_2} \leftrightarrow {B_1}}} + {N_{{A_1} \leftrightarrow {A_2}}}.\ee

\begin{figure}
    \centering
    \includegraphics[scale=0.8]{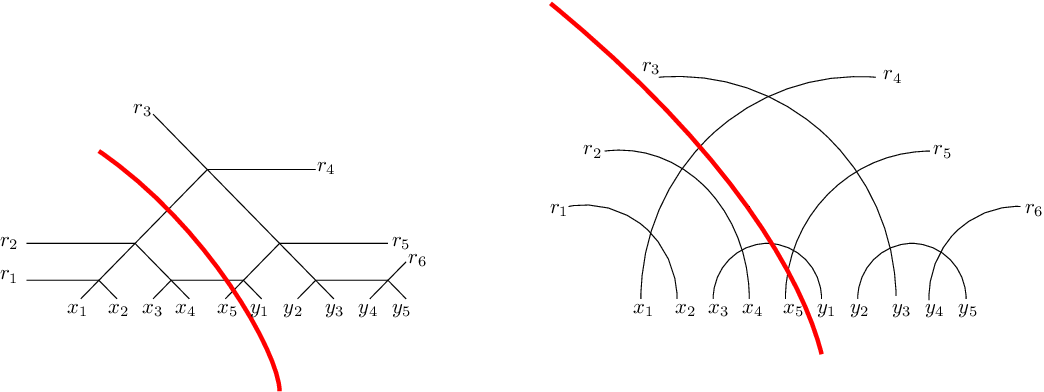}
    \caption{
(a) The entanglement wedge of $A = {A_1} \cup {A_2}$, the red solid line represents EWCS. (b) the minimal entanglement of purification when the coarse-grained state is simply obtained by the direct product of all threads.
}
    \label{fig6.2.2}
\end{figure}

Consider the case where $A = {A_1} \cup {A_2}$ as a whole is a connected region. Recall that in Section~\ref{sec52}, we have obtained a tensor network representation of the reduced density matrix of any connected region, as shown in figure~\ref{fig5.2.10}, which is the adhesion of the entanglement wedge and its mirror image. In this section, we label the sites inside $A_1$ as ${x_i}$, the sites inside $A_2$ as ${y_j}$, and the sites contained in the minimal cut ${\gamma _A}$ of $A$ as ${r_k}$. In particular, we have chosen a concise example as shown in the Fig.\ref{fig6.2.2}, where
\be{A_1} = \{ {x_1},\;{x_2},\;{x_3},\;{x_4},\;{x_5}\} ,\ee
As pointed out in Section \ref{sec52}, the minimal cut ${\gamma _A}$ can be regarded as a purification of the system $A = {A_1} \cup {A_2}$. In other words, the system ${\gamma _A} \equiv \{ {r_k}\} $ can be exactly regarded as the auxiliary system ${A'_1} \cup \;{A'_2}$. The resulting purified state $\psi ({A_1}{A_2}{A'_1}{A'_2})$ is precisely the state $\left| {{\psi _A}} \right\rangle $ corresponding to the entanglement wedge subnetwork of $A$. More importantly, when tracing out $A$ from $\left| {{\psi _A}} \right\rangle $, we obtain that the reduced density matrix of ${\gamma _A}$ is direct-product (see figure~), in other words, it does not contain internal entanglement. Therefore, the auxiliary system ${\gamma _A} = {A'_1} \cup \;{A'_2}$ can now be used to characterize the intrinsic correlation between $A_1$ and $A_2$.

If we regard the thread configuration simply as the direct product of all threads representing bell states, how much is the intrinsic correlation between $A_1$ and $A_2$? For this, we can partition the sites in ${\gamma _A}$ as follows to obtain the minimal entanglement of purification $S({A_1}{A'_1} \leftrightarrow {A_2}{A'_2})$ as a measure of this intrinsic correlation:
\be\label{opt1}{A'_1} = \{ {r_1},\;{r_2},\;{r_4},\;{r_5}\} ,\quad {A'_2} = \{ {r_3},\;{r_6}\}. \ee
How do we know this is the optimal solution? Notice that $x_1$ forms the bell pair with $r_4$ that has the maximum entanglement, and $x_5$ forms the bell pair with $r_5$, and so on. The principle is to try to place pairs of sites with maximum entanglement into the same group. In this way, what truly characterizes the intrinsic correlation between $A_1$ and $A_2$ is only one bell pair, which is represented by the thread connecting $x_3$ and $y_1$. Any other non-optimal purification scheme would introduce redundant correlations. For example, if we factitiously consider sites $r_4$ and $r_5$ as degrees of freedom belonging to ${A'_2}$, it would introduce redundant correlations between ${A'_2}$ and sites $x_1$, $x_5$ in $A_1$. In a word, as implicitly implied in Section \ref{sec32}, when using a coarse-grained state containing only bipartite entanglement, that characterizes the intrinsic correlation between $A_1$ and $A_2$ is only the number ${N_{{A_1} \leftrightarrow {A_2}}}$ of threads directly connecting $A_1$ and $A_2$ (such as $x_3$ connecting to $y_1$) :
\be{E_P}({A_1}:{A_2}) = \mathop {\min }\limits_{{{\left| \psi  \right\rangle }_{{A_1}{{A'}_1}{A_2}{{A'}_2}}}} S({A_1}{A'_1}) = {N_{{A_1} \leftrightarrow {A_2}}},\ee
and this equation contradicts (\ref{equ2}).

\begin{figure}
     \centering
     \begin{subfigure}[b]{1.0\textwidth}
         \centering
         \includegraphics[width=\textwidth]{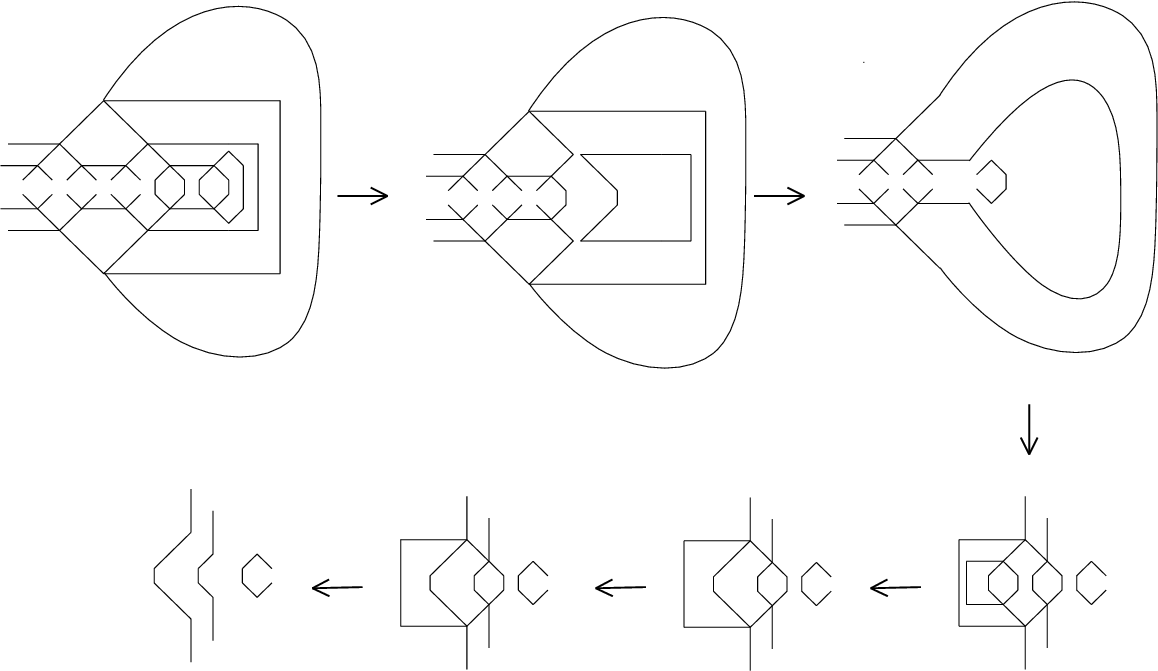}
         \caption{}
         \label{fig6.2.3}
     \end{subfigure}
     \hfill
     \begin{subfigure}[b]{0.7\textwidth}
         \centering
         \includegraphics[width=\textwidth]{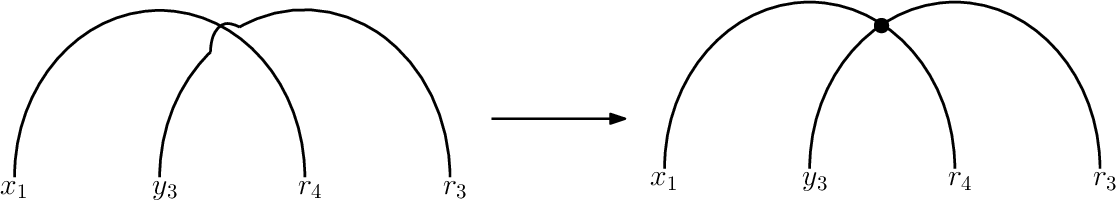}
         \caption{}
         \label{fig6.2.4}
     \end{subfigure}     
            \caption{(a) The calculation of the entanglement entropy $S({A_1}{A'_1} \leftrightarrow {A_2}{A'_2})$ according to the optimal purification scheme expected from the surface-state duality. (b) The introduction of perfect entanglement. }
        \label{}
\end{figure}

This is not a mathematical difficulty but signifies a rigid physical requirement: we must introduce ingredients beyond bipartite entanglement. The perfect entanglement scheme can indeed resolve this physical dilemma. As shown in figure~\ref{fig6.2.3}, let us verify the expected optimal purification scheme from the surface-state duality, i.e.,
\be{A'_1} = \{ {r_1},\;{r_2}\} ,\quad {A'_2} = \{ {r_3},\;{r_4},\;{r_5},\;{r_6}\}. \ee
By this, we perform the calculation of the entanglement entropy $S({A_1}{A'_1} \leftrightarrow {A_2}{A'_2})$, which means tracing:
\be{\rho _{{A_1}{{A'}_1}}} = t{r_{{A_2}{{A'}_2}}}\psi ({A_1}{A_2}{A'_1}{A'_2}).\ee
The graphical algorithm is shown in Fig.\ref{fig6.2.3}. Similarly, at the step~, we can again use the technique of exchanging the trace to calculate the entanglement entropy. The ultimately simplified net result is consisting of three direct-product threads, which gives the correct answer (\ref{ans}). Interestingly, we can also identify from the figure that the sources of these three threads exactly correspond to all the green and purple threads in Fig.\ref{fig6.2.1}. Readers can also verify that if the purification scheme (\ref{opt1}) is adopted, a larger entropy will be obtained.

Although we have presented a detailed calculation, it is not difficult to straightforwardly grasp the most essential nature. Seeing Figure~\ref{fig6.2.4}, focusing on the site ${x_1} \in {A_1}$ and the site ${y_3} \in {A_2}$, before applying perfect entanglement to the coarse-grained state, there is obviously no entanglement between ${x_1}$ and ${y_3}$. Therefore, it cannot contribute to the correlation between $A_1$ and $A_2$. However, after coupling the thread ${\zeta _{{x_1}{r_4}}}$ and the thread ${\zeta _{{y_3}{r_3}}}$ into a perfect tensor state, ${x_1}$ and ${y_3}$ are locally living in a full entangled state. At this point, we must find a way to measure the intrinsic correlation between ${x_1}$ and ${y_3}$. A natural way is entanglement of purification, and the optimal purification scheme is to choose ${r_3} \cup {r_4}$ as the auxiliary system associated with ${y_3}$ and choose the empty set as the auxiliary system associated with ${x_1}$. In this way, the amount of entanglement obtained will be one times log3.

\section{Conclusion and Discussion}

In recent years, many concepts and tools from quantum information theory have been employed to study the quantum entanglement structure in holographic duality. It is extremely challenging to directly derive the mechanism that precisely generates the quantum entanglement structure of the dual spacetime in one fell swoop. Therefore, we choose to explore this entanglement structure at a coarse-grained level, which has generated some insightful and crucial clues. More specifically, we focus on a core concept: conditional mutual information, constructing a class of coarse-grained states intuitively related to a family of thread configurations. It is noteworthy that such coarse-grained states are closely connected to concepts such as kinematic space, holographic entropy cone, holographic partial entanglement entropy, and so on. However, fundamentally, these coarse-grained states are just direct-product states of bipartite entangled states. When we attempt to use these coarse-grained states to further characterize some quantum information theory quantities with geometric duals in holographic duality, such as entanglement entropy of disconnected regions and entanglement of purification dual to entanglement wedge cross section, unavoidable difficulties arise even at the coarse-grained level. On the other hand, introducing perfect tensor entanglement with permutation symmetry naturally solves these problems.

In summary, there are several notable findings in this paper. Firstly, our work, in a sense, provides a equivalency between two quantum information theory quantities of entanglement wedge cross section—entanglement of purification and balanced partial entropy (i.e., expression (\ref{sig})). This relies on our physical interpretation of replacing bipartite entanglement with perfect entanglement. Before this, the expression (\ref{sig}) should be regarded as a noteworthy ``experimental phenomenon". Only when adopting this understanding of perfect entanglement can the expression (\ref{sig}) reasonably characterize the intrinsic correlation between two parts of a system in a mixed state. Secondly, we reexamine, in a sense, the connection between MERA structure and kinematic space. We construct a coarse-grained state with MERA structure, and it is noteworthy that this coarse-grained state does not need to completely characterize the ground state of the holographic CFT, but only delineates its entanglement structure at the coarse-grained level. However, this structure is already sufficient to present the key features of kinematic space: its volume density is precisely given by conditional mutual information. Moreover, our investigation indicates that the spatial points in kinematic space should not be viewed as being direct-product but, in some sense, entangled together by perfect entanglement. Additionally, the thread configurations with MERA structure that we construct have an insightful role in the relation between MERA tensor networks and bit threads. In this paper, we have not considered the context of BTZ black holes~\cite{Banados:1992wn,Vidal:2015,Czech:2015xna,Gan:2016vuw}. A very natural idea is to further consider the thread configurations dual to the MERA structure of dual BTZ black holes and their corresponding coarse-grained states. What is more intriguing is that wormhole geometry has been argued to be closely related to entanglement wedge cross section\cite{Bao:2018fso,Bao:2018zab}, stemming from the fact that BTZ black holes can be viewed as quotients of AdS space. Thus, we expect to once again see the necessity of perfect entanglement in the context of BTZ black holes, which may appear in a more nontrivial way. We leave this fascinating idea for future work.

\begin{appendix}
\section*{Appendix}

This paper involves a significant amount of tensor diagram calculus. Therefore, here are some brief rules for computing with tensor networks.

\begin{figure}
     \centering
     \begin{subfigure}[b]{0.3\textwidth}
         \centering
         \includegraphics[width=\textwidth]{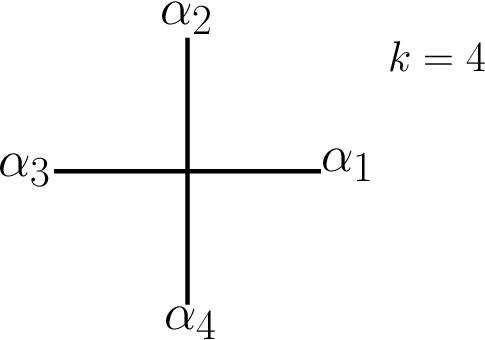}
         \caption{}
         \label{figappendix1a}
     \end{subfigure}
     \hspace{13em}
     \begin{subfigure}[b]{0.3\textwidth}
         \centering
         \includegraphics[width=\textwidth]{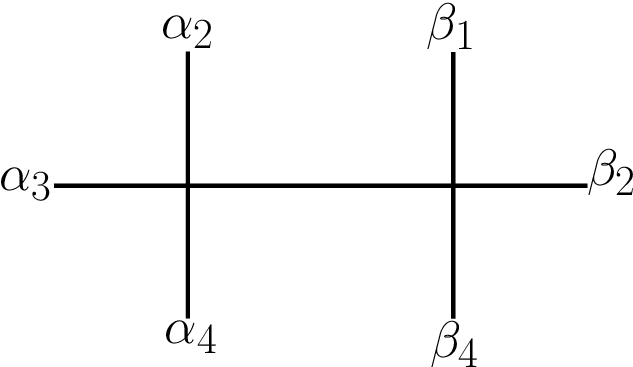}
         \caption{}
         \label{figappendix1b}
     \end{subfigure}
   \hspace{13em}
     \begin{subfigure}[b]{0.2\textwidth}
         \centering
         \includegraphics[width=\textwidth]{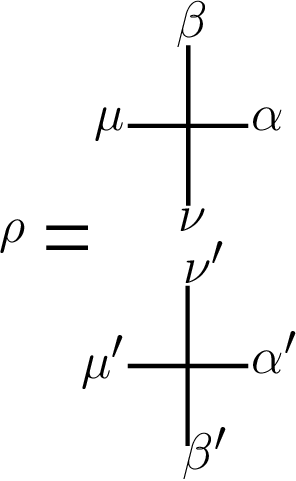}
         \caption{}
         \label{figappendix2a}
     \end{subfigure}
     \hspace{13em}
     \begin{subfigure}[b]{0.3\textwidth}
         \centering
         \includegraphics[width=\textwidth]{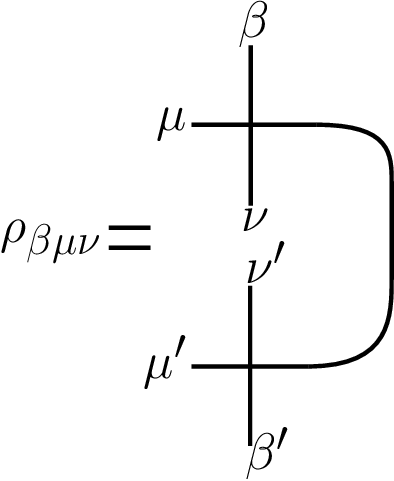}
         \caption{}
         \label{figappendix2b}
     \end{subfigure}
      \hspace{26em}
      
     \begin{subfigure}[b]{0.9\textwidth}
         \centering
         \includegraphics[width=\textwidth]{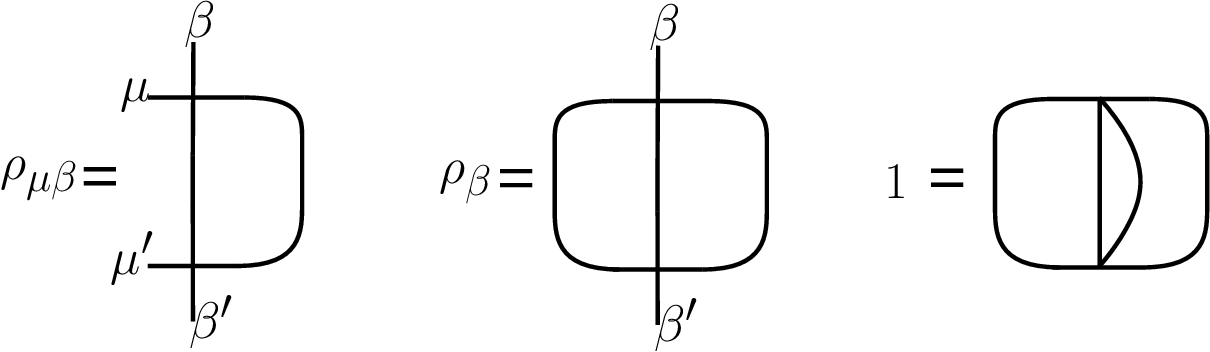}
         \caption{}
         \label{figappendix3}
     \end{subfigure}
     \caption{Basic rules for tensor diagram calculus.}
    \label{appendix1}
\end{figure}

We can represent a normalized quantum state over k sites (or qudits) as a simple diagram T, which consists of k legs extending from a vertex:
\be\label{ki}\left| \chi \right\rangle = {T_{{\alpha_1}{\alpha_2} \cdots {\alpha_k}}}\left| {{\alpha_1}{\alpha_2} \cdots {\alpha_k}} \right\rangle. \ee
In Fig.~\ref{figappendix1a}, we depict the case of k=4, which is the primary case used in this paper. Accordingly, each endpoint ${\alpha_i}$ of the legs of T represents a qudit equipped with a Hilbert space of dimension ${d_i}$. With this, the tensor contraction operation can be translated into diagrammatic calculations involving T. For example, consider the operation:
\be\left| \psi \right\rangle = {T_{{\alpha_1}{\alpha_2}{\alpha_3}{\alpha_4}}}{T_{{\beta_1}{\alpha_2}{\beta_3}{\beta_4}}}\left| {{\alpha_1}{\alpha_3}{\alpha_4}{\beta_1}{\beta_3}{\beta_4}} \right\rangle ,\ee
its graphical representation is shown in Fig.~\ref{figappendix1b}, where the index contractions are represented by two corresponding legs being glued together into an inner leg. Thus, by continually gluing basic patterns, one can obtain more complex pure states of a large number of qudits.

The above procedure can be extended to operations involving density matrices. Taking the example of the state (\ref{ki}) for k=4, its corresponding density matrix, shown in Fig.~\ref{figappendix2a}, is:
\be\rho = {T_{\alpha \beta \mu \nu }}{T_{\alpha '\beta '\mu '\nu '}}\left| {\alpha \beta \mu \nu } \right\rangle \left\langle {\alpha '\beta '\mu '\nu '} \right|.\ee
Due to the apparent symmetry, we represent it as the original tensor network representing a pure state accompanied by its mirror image. Additionally, to distinguish it from the representation of the pure state, we arrange it in an upper-lower configuration.

Now, what is interesting is that the operation of taking the trace of the density matrix becomes a diagrammatic calculation. For instance, the reduced density matrix:
\be{\rho _{\beta \mu \nu }} = t{r_\alpha }\rho = \sum\limits_{\alpha ''} {\left\langle {\alpha ''} \right|\rho \left| {\alpha ''} \right\rangle } = {T_{\alpha ''\beta \mu \nu }}{T_{\alpha ''\beta '\mu '\nu '}}\left| {\beta \mu \nu } \right\rangle \left\langle {\beta '\mu '\nu '} \right|,\ee
corresponds to Fig.~\ref{figappendix2b}.

Similarly, we can obtain the following examples as shown in diagram~\ref{figappendix3}:
\be{\rho _{\mu \beta }} = t{r_{\alpha \nu }}\rho ,\ee
\be{\rho _\beta } = t{r_{\alpha \mu \nu }}\rho ,\ee
and especially,
\be t{r_{\alpha \beta \mu \nu }}\rho = 1.\ee

\end{appendix}

\newpage{\pagestyle{empty}\cleardoublepage}


\begin{thebibliography}{100}

	\bibitem{Maldacena:1997re} J.~M.~Maldacena,
``The Large N limit of superconformal field theories and supergravity,''
Adv. Theor. Math. Phys. \textbf{2}, 231-252 (1998)
[arXiv:hep-th/9711200 [hep-th]].


	\bibitem{Gubser:1998bc} S.~S.~Gubser, I.~R.~Klebanov and A.~M.~Polyakov,
``Gauge theory correlators from noncritical string theory,''
Phys. Lett. B \textbf{428}, 105-114 (1998)
[arXiv:hep-th/9802109 [hep-th]].


	\bibitem{Witten:1998qj} E.~Witten,
``Anti-de Sitter space and holography,''
Adv. Theor. Math. Phys. \textbf{2}, 253-291 (1998)
[arXiv:hep-th/9802150 [hep-th]].


	\bibitem{Ryu:2006bv} S.~Ryu and T.~Takayanagi,
``Holographic derivation of entanglement entropy from AdS/CFT,''
Phys. Rev. Lett. \textbf{96}, 181602 (2006)
[arXiv:hep-th/0603001 [hep-th]].


	\bibitem{Ryu:2006ef} S.~Ryu and T.~Takayanagi,
``Aspects of Holographic Entanglement Entropy,''
JHEP \textbf{08}, 045 (2006)
[arXiv:hep-th/0605073 [hep-th]].


	\bibitem{Hubeny:2007xt} V.~E.~Hubeny, M.~Rangamani and T.~Takayanagi,
``A Covariant holographic entanglement entropy proposal,''
JHEP \textbf{07}, 062 (2007)
[arXiv:0705.0016 [hep-th]].

  
    
  
    
	

	\bibitem{Vidal:2007hda} G.~Vidal,
	``Entanglement Renormalization,''
	Phys. Rev. Lett. \textbf{99}, no.22, 220405 (2007)
	[arXiv:cond-mat/0512165 [cond-mat]].
	

	\bibitem{Vidal:2008zz} G.~Vidal,
	``Class of Quantum Many-Body States That Can Be Efficiently Simulated,''
	Phys. Rev. Lett. \textbf{101}, 110501 (2008)
	[arXiv:quant-ph/0610099 [quant-ph]].


	\bibitem{Vidal:2015} E.~Glen, G.~Vidal,
 ``Tensor network renormalization yields the multiscale entanglement renormalization ansatz,''
 Phys. Rev. Lett. \textbf{115},200401 (2015).
	

	\bibitem{Swingle:2009bg} B.~Swingle,
	``Entanglement Renormalization and Holography,''
	Phys. Rev. D \textbf{86}, 065007 (2012)
	[arXiv:0905.1317 [cond-mat.str-el]].
	

	\bibitem{Swingle:2012wq} B.~Swingle,
	``Constructing holographic spacetimes using entanglement renormalization,''
	[arXiv:1209.3304 [hep-th]].
	

	\bibitem{Pastawski:2015qua} F.~Pastawski, B.~Yoshida, D.~Harlow and J.~Preskill,
	``Holographic quantum error-correcting codes: Toy models for the bulk/boundary correspondence,''
	JHEP \textbf{06}, 149 (2015)
	[arXiv:1503.06237 [hep-th]].
	

	\bibitem{Hayden:2016cfa} P.~Hayden, S.~Nezami, X.~L.~Qi, N.~Thomas, M.~Walter and Z.~Yang,
	``Holographic duality from random tensor networks,''
	JHEP \textbf{11}, 009 (2016)
	[arXiv:1601.01694 [hep-th]].
	

	\bibitem{Chen:2021ipv} L.~Chen, X.~Liu and L.~Y.~Hung,
	``Emergent Einstein Equation in p-adic Conformal Field Theory Tensor Networks,''
	Phys. Rev. Lett. \textbf{127}, no.22, 221602 (2021)
	[arXiv:2102.12022 [hep-th]].
	

	\bibitem{Chen:2021qah} L.~Chen, X.~Liu and L.~Y.~Hung,
	``Bending the Bruhat-Tits tree. Part II. The p-adic BTZ black hole and local diffeomorphism on the Bruhat-Tits tree,''
	JHEP \textbf{09}, 097 (2021)
	[arXiv:2102.12024 [hep-th]].
	

	\bibitem{Chen:2021rsy} L.~Chen, X.~Liu and L.~Y.~Hung,
	``Bending the Bruhat-Tits tree. Part I. Tensor network and emergent Einstein equations,''
	JHEP \textbf{06}, 094 (2021)
	[arXiv:2102.12023 [hep-th]].
	

	\bibitem{Bao:2018pvs} N.~Bao, G.~Penington, J.~Sorce and A.~C.~Wall,
	``Beyond Toy Models: Distilling Tensor Networks in Full AdS/CFT,''
	JHEP \textbf{11}, 069 (2019)
	[arXiv:1812.01171 [hep-th]].
	

	\bibitem{Bao:2019fpq} N.~Bao, G.~Penington, J.~Sorce and A.~C.~Wall,
	``Holographic Tensor Networks in Full AdS/CFT,''
	[arXiv:1902.10157 [hep-th]].
	

	\bibitem{Haegeman:2011uy} J.~Haegeman, T.~J.~Osborne, H.~Verschelde and F.~Verstraete,
	``Entanglement Renormalization for Quantum Fields in Real Space,''
	Phys. Rev. Lett. \textbf{110}, no.10, 100402 (2013)
	[arXiv:1102.5524 [hep-th]].
    

	

	\bibitem{Qi:2013caa} X.~L.~Qi,
   ``Exact holographic mapping and emergent space-time geometry,''
   [arXiv:1309.6282 [hep-th]].


	\bibitem{Miyaji:2015fia} M.~Miyaji, T.~Numasawa, N.~Shiba, T.~Takayanagi and K.~Watanabe,
	``Continuous Multiscale Entanglement Renormalization Ansatz as Holographic Surface-State Correspondence,''
	Phys. Rev. Lett. \textbf{115}, no.17, 171602 (2015)
	[arXiv:1506.01353 [hep-th]].


	\bibitem{Miyaji:2015yva} M.~Miyaji and T.~Takayanagi,
	``Surface/State Correspondence as a Generalized Holography,''
	PTEP \textbf{2015}, no.7, 073B03 (2015)
	[arXiv:1503.03542 [hep-th]].
	

	\bibitem{Chen:2022wvy} L.~Chen, H.~Zhang, H.~C.~Zhang, K.~X.~Ji, C.~Shen, R.~s.~Wang, X.~d.~Zeng and L.~Y.~Hung,
``Exact Holographic Tensor Networks -- Constructing CFT$_D$ from TQFT$_{D+1}$,''
[arXiv:2210.12127 [hep-th]].


	\bibitem{Almheiri:2014lwa} A.~Almheiri, X.~Dong and D.~Harlow,
``Bulk Locality and Quantum Error Correction in AdS/CFT,''
JHEP \textbf{04}, 163 (2015)
[arXiv:1411.7041 [hep-th]].


    
    

	\bibitem{Czech:2015xna} B.~Czech, G.~Evenbly, L.~Lamprou, S.~McCandlish, X.~L.~Qi, J.~Sully and G.~Vidal,
``Tensor network quotient takes the vacuum to the thermal state,''
Phys. Rev. B \textbf{94}, no.8, 085101 (2016)
[arXiv:1510.07637 [cond-mat.str-el]].


	\bibitem{Evenbly:2017hyg} G.~Evenbly,
``Hyperinvariant Tensor Networks and Holography,''
Phys. Rev. Lett. \textbf{119}, no.14, 141602 (2017)
[arXiv:1704.04229 [quant-ph]].
    

	\bibitem{Steinberg:2023wll} M.~Steinberg, S.~Feld and A.~Jahn,
``Holographic codes from hyperinvariant tensor networks,''
Nature Commun. \textbf{14}, no.1, 7314 (2023)

[arXiv:2304.02732 [quant-ph]].


	\bibitem{Steinberg:2020bef} M.~Steinberg and J.~Prior,
``Conformal properties of hyperinvariant tensor networks,''
Sci. Rep. \textbf{12}, no.1, 532 (2022)
[arXiv:2012.09591 [quant-ph]].


	\bibitem{Czech:2016nxc} B.~Czech, P.~H.~Nguyen and S.~Swaminathan,
``A defect in holographic interpretations of tensor networks,''
JHEP \textbf{03}, 090 (2017)
[arXiv:1612.05698 [hep-th]].


	\bibitem{Singh:2017tet} S.~Singh, N.~A.~McMahon and G.~K.~Brennen,
``Holographic spin networks from tensor network states,''
Phys. Rev. D \textbf{97}, no.2, 026013 (2018)
[arXiv:1702.00392 [cond-mat.str-el]].



	\bibitem{Colafranceschi:2022ual} E.~Colafranceschi and G.~Adesso,
``Holographic entanglement in spin network states: A focused review,''
AVS Quantum Sci. \textbf{4}, no.2, 025901 (2022)
[arXiv:2202.05116 [hep-th]].
   
    

	\bibitem{Cheng:2023kxh} G.~Cheng, L.~Chen, Z.~C.~Gu and L.~Y.~Hung,
``Exact fixed-point tensor network construction for rational conformal field theory,''
[arXiv:2311.18005 [cond-mat.str-el]].


	\bibitem{Zeng:2023dzh} X.~Zeng and L.~Y.~Hung,
``Bulk Operator Reconstruction in Topological Tensor Network and Generalized Free Fields,''
Entropy \textbf{25}, no.11, 1543 (2023)
[arXiv:2309.03178 [hep-th]].
   

	\bibitem{Hung:2019zsk} L.~Y.~Hung, W.~Li and C.~M.~Melby-Thompson,
    ``$p$-adic CFT is a holographic tensor network,''
    JHEP \textbf{04}, 170 (2019)
    [arXiv:1902.01411 [hep-th]].
    

	\bibitem{Milsted:2018vop} A.~Milsted and G.~Vidal,
    ``Tensor networks as conformal transformations,''
    [arXiv:1805.12524 [cond-mat.str-el]].
    

	\bibitem{Milsted:2018yur} A.~Milsted and G.~Vidal,
    ``Tensor networks as path integral geometry,''
    [arXiv:1807.02501 [cond-mat.str-el]].
    

	\bibitem{Milsted:2018san} A.~Milsted and G.~Vidal,
    ``Geometric interpretation of the multi-scale entanglement renormalization ansatz,''
    [arXiv:1812.00529 [hep-th]].
    

	\bibitem{SinaiKunkolienkar:2016lgg} R.~Sinai Kunkolienkar and K.~Banerjee,
    ``Towards a dS/MERA correspondence,''
    Int. J. Mod. Phys. D \textbf{26}, no.13, 1750143 (2017)
    [arXiv:1611.08581 [hep-th]].
    

	\bibitem{Bao:2017qmt} N.~Bao, C.~Cao, S.~M.~Carroll and A.~Chatwin-Davies,
    ``De Sitter Space as a Tensor Network: Cosmic No-Hair, Complementarity, and Complexity,''
    Phys. Rev. D \textbf{96}, no.12, 123536 (2017)
    [arXiv:1709.03513 [hep-th]].
    

	\bibitem{Beny:2011vh} C.~Beny,
    ``Causal structure of the entanglement renormalization ansatz,''
    New J. Phys. \textbf{15}, 023020 (2013)
    [arXiv:1110.4872 [quant-ph]].
    

	\bibitem{Czech:2015kbp} B.~Czech, L.~Lamprou, S.~McCandlish and J.~Sully,
    ``Tensor Networks from Kinematic Space,''
    JHEP \textbf{07}, 100 (2016)
    [arXiv:1512.01548 [hep-th]].


	\bibitem{Ling:2019akz} Y.~Ling, Y.~Xiao and M.~H.~Wu,
    ``Note on quantum entanglement and quantum geometry,''
    Phys. Lett. B \textbf{798}, 135023 (2019)
    [arXiv:1907.01215 [hep-th]].
    

	\bibitem{Ling:2018ajv} Y.~Ling, Y.~Liu, Z.~Y.~Xian and Y.~Xiao,
    ``Tensor chain and constraints in tensor networks,''
    JHEP \textbf{06}, 032 (2019)
    [arXiv:1807.10247 [hep-th]].
    

	\bibitem{Ling:2018vza} Y.~Ling, Y.~Liu, Z.~Y.~Xian and Y.~Xiao,
    ``Quantum error correction and entanglement spectrum in tensor networks,''
    Phys. Rev. D \textbf{99}, no.2, 026008 (2019)
    [arXiv:1806.05007 [hep-th]].
    
    

	\bibitem{Bhattacharyya:2017aly} A.~Bhattacharyya, L.~Y.~Hung, Y.~Lei and W.~Li,
   ``Tensor network and ($p$-adic) AdS/CFT,''
   JHEP \textbf{01}, 139 (2018)
   [arXiv:1703.05445 [hep-th]].
   

	\bibitem{Bhattacharyya:2016hbx} A.~Bhattacharyya, Z.~S.~Gao, L.~Y.~Hung and S.~N.~Liu,
    ``Exploring the Tensor Networks/AdS Correspondence,''
    JHEP \textbf{08}, 086 (2016)
    [arXiv:1606.00621 [hep-th]].
    

	\bibitem{Gan:2017nyt} W.~C.~Gan and F.~W.~Shu,
    ``Holography as deep learning,''
    Int. J. Mod. Phys. D \textbf{26}, no.12, 1743020 (2017)
    [arXiv:1705.05750 [gr-qc]].


	\bibitem{Bao:2015uaa} N.~Bao, C.~Cao, S.~M.~Carroll, A.~Chatwin-Davies, N.~Hunter-Jones, J.~Pollack and G.~N.~Remmen,
    ``Consistency conditions for an AdS multiscale entanglement renormalization ansatz correspondence,''
    Phys. Rev. D \textbf{91}, no.12, 125036 (2015)
    [arXiv:1504.06632 [hep-th]].
    

	\bibitem{Yu:2020zwk} C.~Yu, F.~Z.~Chen, Y.~Y.~Lin, J.~R.~Sun and Y.~Sun,
    ``Note on surface growth approach for bulk reconstruction ,''
    Chin. Phys. C \textbf{46}, no.8, 085104 (2022)
    [arXiv:2010.03167 [hep-th]].
    

	\bibitem{Sun:2019ycv} J.~R.~Sun and Y.~Sun,
    ``On the emergence of gravitational dynamics from tensor networks,''
    [arXiv:1912.02070 [hep-th]].
   
    

	\bibitem{Belin:2023efa} A.~Belin, J.~de Boer, D.~L.~Jafferis, P.~Nayak and J.~Sonner,
``Approximate CFTs and Random Tensor Models,''
[arXiv:2308.03829 [hep-th]].

	

	\bibitem{Lin:2020ufd} Y.~Y.~Lin, J.~R.~Sun and Y.~Sun,
	``Surface growth scheme for bulk reconstruction and tensor network,''
	JHEP \textbf{12}, 083 (2020)
	[arXiv:2010.01907 [hep-th]].
	

	\bibitem{Freedman:2016zud} M.~Freedman and M.~Headrick,
	``Bit threads and holographic entanglement,''
	Commun. Math. Phys. \textbf{352}, no.1, 407-438 (2017)
	[arXiv:1604.00354 [hep-th]].
	

	\bibitem{Cui:2018dyq} S.~X.~Cui, P.~Hayden, T.~He, M.~Headrick, B.~Stoica and M.~Walter,
	``Bit Threads and Holographic Monogamy,''
	Commun. Math. Phys. \textbf{376}, no.1, 609-648 (2019)
	[arXiv:1808.05234 [hep-th]].
	

	\bibitem{Headrick:2017ucz} M.~Headrick and V.~E.~Hubeny,
	``Riemannian and Lorentzian flow-cut theorems,''
	Class. Quant. Grav. \textbf{35}, no.10, 10 (2018)
	[arXiv:1710.09516 [hep-th]].


	\bibitem{Headrick:2022nbe} M.~Headrick and V.~E.~Hubeny,
``Covariant bit threads,''
JHEP \textbf{07}, 180 (2023)
[arXiv:2208.10507 [hep-th]].



	\bibitem{Headrick:2020gyq} M.~Headrick, J.~Held and J.~Herman,
``Crossing Versus Locking: Bit Threads and Continuum Multiflows,''
Commun. Math. Phys. \textbf{396}, no.1, 265-313 (2022)
[arXiv:2008.03197 [hep-th]].
	

	\bibitem{Lin:2020yzf} Y.~Y.~Lin, J.~R.~Sun and Y.~Sun,
	``Bit thread, entanglement distillation, and entanglement of purification,''
	Phys. Rev. D \textbf{103}, no.12, 126002 (2021)
	[arXiv:2012.05737 [hep-th]].
	

	\bibitem{Lin:2021hqs} Y.~Y.~Lin, J.~R.~Sun and J.~Zhang,
	``Deriving the PEE proposal from the locking bit thread configuration,''
	JHEP \textbf{10}, 164 (2021)

	[arXiv:2105.09176 [hep-th]].
	


	\bibitem{Lin:2022aqf} Y.~Y.~Lin, J.~R.~Sun, Y.~Sun and J.~C.~Jin,
    ``The PEE aspects of entanglement islands from bit threads,''

    JHEP \textbf{07}, 009 (2022)
    [arXiv:2203.03111 [hep-th]].



	\bibitem{Lin:2022flo} Y.~Y.~Lin and J.~C.~Jin,
``Thread/State correspondence: from bit threads to qubit threads,''
JHEP \textbf{02}, 245 (2023)
[arXiv:2210.08783 [hep-th]].


	\bibitem{Lin:2022agc} Y.~Y.~Lin and J.~C.~Jin,
``Thread/State correspondence: the qubit threads model of holographic gravity,''
[arXiv:2208.08963 [hep-th]].


	\bibitem{Lin:2023orb} Y.~Y.~Lin,
``Distilled density matrices of holographic partial entanglement entropy from thread-state correspondence,''
Phys. Rev. D \textbf{108}, no.10, 106010 (2023)
[arXiv:2305.02895 [hep-th]].


    

	\bibitem{Kudler-Flam:2019oru} J.~Kudler-Flam, I.~MacCormack and S.~Ryu,
``Holographic entanglement contour, bit threads, and the entanglement tsunami,''
J. Phys. A \textbf{52}, no.32, 325401 (2019)
[arXiv:1902.04654 [hep-th]].


	\bibitem{Rolph:2021nan} A.~Rolph,
``Local measures of entanglement in black holes and CFTs,''
SciPost Phys. \textbf{12}, no.3, 079 (2022)
[arXiv:2107.11385 [hep-th]].

    

	\bibitem{Harper:2022sky} J.~Harper,
``Perfect tensor hyperthreads,''
JHEP \textbf{09}, 239 (2022)
[arXiv:2205.01140 [hep-th]].


	\bibitem{Harper:2021uuq} J.~Harper,
``Hyperthreads in holographic spacetimes,''
JHEP \textbf{09}, 118 (2021)
[arXiv:2107.10276 [hep-th]].


	\bibitem{Lin:2023rbd} J.~Lin, Y.~Lu and Q.~Wen,
``Geometrizing the Partial Entanglement Entropy: from PEE Threads to Bit Threads,''
[arXiv:2311.02301 [hep-th]].
    

	\bibitem{Agon:2021tia} C.~A.~Ag\'on and J.~F.~Pedraza,
    ``Quantum bit threads and holographic entanglement,''
    JHEP \textbf{02}, 180 (2022)
    [arXiv:2105.08063 [hep-th]].

    

	\bibitem{Rolph:2021hgz} A.~Rolph,
    ``Quantum bit threads,''
    [arXiv:2105.08072 [hep-th]].
    

	\bibitem{Chen:2018ywy} C.~B.~Chen, F.~W.~Shu and M.~H.~Wu,
    ``Quantum bit threads of MERA tensor network in large $c$ limit,''
    Chin. Phys. C \textbf{44}, no.7, 075102 (2020)
    [arXiv:1804.00441 [hep-th]].
    

	\bibitem{Hubeny:2018bri} V.~E.~Hubeny,
    ``Bulk locality and cooperative flows,''
    JHEP \textbf{12}, 068 (2018)
    [arXiv:1808.05313 [hep-th]].
    

	\bibitem{Agon:2018lwq} C.~A.~Ag\'on, J.~De Boer and J.~F.~Pedraza,
    ``Geometric Aspects of Holographic Bit Threads,''
    JHEP \textbf{05}, 075 (2019)
    [arXiv:1811.08879 [hep-th]].
    

	\bibitem{Du:2019emy} D.~H.~Du, C.~B.~Chen and F.~W.~Shu,
    ``Bit threads and holographic entanglement of purification,''
    JHEP \textbf{08}, 140 (2019)
    [arXiv:1904.06871 [hep-th]].
    

	\bibitem{Bao:2019wcf} N.~Bao, A.~Chatwin-Davies, J.~Pollack and G.~N.~Remmen,
    ``Towards a Bit Threads Derivation of Holographic Entanglement of Purification,''
    JHEP \textbf{07}, 152 (2019)
    [arXiv:1905.04317 [hep-th]].
    

	\bibitem{Harper:2019lff} J.~Harper and M.~Headrick,
    ``Bit threads and holographic entanglement of purification,''
    JHEP \textbf{08}, 101 (2019)
    [arXiv:1906.05970 [hep-th]].
    

	\bibitem{Agon:2019qgh} C.~A.~Ag\'on and M.~Mezei,
    ``Bit threads and the membrane theory of entanglement dynamics,''
    JHEP \textbf{11}, 167 (2021)
    [arXiv:1910.12909 [hep-th]].
    

	\bibitem{Du:2019vwh} D.~H.~Du, F.~W.~Shu and K.~X.~Zhu,
    ``Inequalities of Holographic Entanglement of Purification from Bit Threads,''
    Eur. Phys. J. C \textbf{80}, no.8, 700 (2020)
    [arXiv:1912.00557 [hep-th]].

    

	\bibitem{Agon:2020mvu} C.~A.~Ag\'on, E.~C\'aceres and J.~F.~Pedraza,
    ``Bit threads, Einstein\textquoteright{}s equations and bulk locality,''
    JHEP \textbf{01}, 193 (2021)
    [arXiv:2007.07907 [hep-th]].
    

	\bibitem{Bao:2020uku} N.~Bao and J.~Harper,
    ``Bit threads on hypergraphs,''
    [arXiv:2012.07872 [hep-th]].

    

	\bibitem{Pedraza:2021fgp} J.~F.~Pedraza, A.~Russo, A.~Svesko and Z.~Weller-Davies,
    ``Sewing spacetime with Lorentzian threads: complexity and the emergence of time in quantum gravity,''
    JHEP \textbf{02}, 093 (2022)
    [arXiv:2106.12585 [hep-th]].
    

	\bibitem{Pedraza:2021mkh} J.~F.~Pedraza, A.~Russo, A.~Svesko and Z.~Weller-Davies,
    ``Lorentzian Threads as Gatelines and Holographic Complexity,''
    Phys. Rev. Lett. \textbf{127}, no.27, 271602 (2021)
    [arXiv:2105.12735 [hep-th]].
    

	\bibitem{Harper:2018sdd} J.~Harper, M.~Headrick and A.~Rolph,
    ``Bit Threads in Higher Curvature Gravity,''
    JHEP \textbf{11}, 168 (2018)
    [arXiv:1807.04294 [hep-th]].
    

	\bibitem{Shaghoulian:2022fop} E.~Shaghoulian and L.~Susskind,
    ``Entanglement in De Sitter Space,''
    [arXiv:2201.03603 [hep-th]].
    

	\bibitem{Susskind:2021esx} L.~Susskind,
    ``Entanglement and Chaos in De Sitter Space Holography: An SYK Example,''
    JHAP \textbf{1}, no.1, 1-22 (2021)
    [arXiv:2109.14104 [hep-th]].

    	\bibitem{Bakhmatov:2017ihw} I.~Bakhmatov, N.~S.~Deger, J.~Gutowski, E.~\'O.~Colg\'ain and H.~Yavartanoo,
    ``Calibrated Entanglement Entropy,''
    JHEP \textbf{07}, 117 (2017)
    [arXiv:1705.08319 [hep-th]].
    

    
    

	\bibitem{Czech:2015qta} B.~Czech, L.~Lamprou, S.~McCandlish and J.~Sully,
``Integral Geometry and Holography,''
JHEP \textbf{10}, 175 (2015)
[arXiv:1505.05515 [hep-th]].


	\bibitem{Bao:2015bfa} N.~Bao, S.~Nezami, H.~Ooguri, B.~Stoica, J.~Sully and M.~Walter,
``The Holographic Entropy Cone,''
JHEP \textbf{09}, 130 (2015)
[arXiv:1505.07839 [hep-th]].

    

	\bibitem{Hubeny:2018ijt} V.~E.~Hubeny, M.~Rangamani and M.~Rota,
``The holographic entropy arrangement,''
Fortsch. Phys. \textbf{67}, no.4, 1900011 (2019)
[arXiv:1812.08133 [hep-th]].
	

	\bibitem{Hubeny:2018trv} V.~E.~Hubeny, M.~Rangamani and M.~Rota,
``Holographic entropy relations,''
Fortsch. Phys. \textbf{66}, no.11-12, 1800067 (2018)
[arXiv:1808.07871 [hep-th]].


	\bibitem{HernandezCuenca:2019wgh} S.~Hern\'andez Cuenca,
``Holographic entropy cone for five regions,''
Phys. Rev. D \textbf{100}, no.2, 026004 (2019)
[arXiv:1903.09148 [hep-th]].
		
	




	\bibitem{Vidal:2014aal} G.~Vidal and Y.~Chen,
``Entanglement contour,''
J. Stat. Mech. \textbf{2014}, no.10, P10011 (2014)
[arXiv:1406.1471 [cond-mat.str-el]].


	\bibitem{Wen:2019iyq} Q.~Wen,
``Formulas for Partial Entanglement Entropy,''
Phys. Rev. Res. \textbf{2}, no.2, 023170 (2020)
[arXiv:1910.10978 [hep-th]].


	\bibitem{Wen:2018whg} Q.~Wen,
``Fine structure in holographic entanglement and entanglement contour,''
Phys. Rev. D \textbf{98}, no.10, 106004 (2018)
[arXiv:1803.05552 [hep-th]].



	\bibitem{Wen:2020ech} Q.~Wen,
``Entanglement contour and modular flow from subset entanglement entropies,''
JHEP \textbf{05}, 018 (2020)
[arXiv:1902.06905 [hep-th]].


	\bibitem{Han:2019scu} M.~Han and Q.~Wen,
``Entanglement entropy from entanglement contour: higher dimensions,''
SciPost Phys. Core \textbf{5}, 020 (2022)
[arXiv:1905.05522 [hep-th]].





	\bibitem{Nguyen:2017yqw} P.~Nguyen, T.~Devakul, M.~G.~Halbasch, M.~P.~Zaletel and B.~Swingle,
``Entanglement of purification: from spin chains to holography,''
JHEP \textbf{01}, 098 (2018)
[arXiv:1709.07424 [hep-th]].


	\bibitem{Takayanagi:2017knl} T.~Takayanagi and K.~Umemoto,
``Entanglement of purification through holographic duality,''
Nature Phys. \textbf{14}, no.6, 573-577 (2018)
[arXiv:1708.09393 [hep-th]].


	\bibitem{Wen:2021qgx} Q.~Wen,
``Balanced Partial Entanglement and the Entanglement Wedge Cross Section,''
JHEP \textbf{04}, 301 (2021)
[arXiv:2103.00415 [hep-th]].


	\bibitem{Camargo:2022mme} H.~A.~Camargo, P.~Nandy, Q.~Wen and H.~Zhong,
``Balanced partial entanglement and mixed state correlations,''
SciPost Phys. \textbf{12}, no.4, 137 (2022)
[arXiv:2201.13362 [hep-th]].


	\bibitem{Wen:2022jxr} Q.~Wen and H.~Zhong,
``Covariant entanglement wedge cross-section, balanced partial entanglement and gravitational anomalies,''
SciPost Phys. \textbf{13}, no.3, 056 (2022)
[arXiv:2205.10858 [hep-th]].




	\bibitem{Helwig:2012nha} W.~Helwig, W.~Cui, A.~Riera, J.~I.~Latorre and H.~K.~Lo,
``Absolute Maximal Entanglement and Quantum Secret Sharing,''
Phys. Rev. A \textbf{86}, 052335 (2012)
[arXiv:1204.2289 [quant-ph]].


	\bibitem{Helwig:2013qoq} W.~Helwig,
``Absolutely Maximally Entangled Qudit Graph States,''
[arXiv:1306.2879 [quant-ph]].




	\bibitem{Dutta:2019gen} S.~Dutta and T.~Faulkner,
``A canonical purification for the entanglement wedge cross-section,''
JHEP \textbf{03}, 178 (2021)
[arXiv:1905.00577 [hep-th]].


	\bibitem{Kudler-Flam:2018qjo} J.~Kudler-Flam and S.~Ryu,
``Entanglement negativity and minimal entanglement wedge cross sections in holographic theories,''
Phys. Rev. D \textbf{99}, no.10, 106014 (2019)
[arXiv:1808.00446 [hep-th]].



	\bibitem{Kusuki:2019zsp} Y.~Kusuki, J.~Kudler-Flam and S.~Ryu,
``Derivation of holographic negativity in AdS$_3$/CFT$_2$,''
Phys. Rev. Lett. \textbf{123}, no.13, 131603 (2019)
[arXiv:1907.07824 [hep-th]].


	\bibitem{Tamaoka:2018ned} K.~Tamaoka,
``Entanglement Wedge Cross Section from the Dual Density Matrix,''
Phys. Rev. Lett. \textbf{122}, no.14, 141601 (2019)
[arXiv:1809.09109 [hep-th]].


	\bibitem{Espindola:2018ozt} R.~Esp\'\i{}ndola, A.~Guijosa and J.~F.~Pedraza,
``Entanglement Wedge Reconstruction and Entanglement of Purification,''
Eur. Phys. J. C \textbf{78}, no.8, 646 (2018)
[arXiv:1804.05855 [hep-th]].




	\bibitem{Nozaki:2013wia} M.~Nozaki, T.~Numasawa and T.~Takayanagi,
``Holographic Local Quenches and Entanglement Density,''
JHEP \textbf{05}, 080 (2013)
[arXiv:1302.5703 [hep-th]].
    

	\bibitem{Bao:2018fso} N.~Bao, A.~Chatwin-Davies and G.~N.~Remmen,
``Entanglement of Purification and Multiboundary Wormhole Geometries,''
JHEP \textbf{02}, 110 (2019)
[arXiv:1811.01983 [hep-th]].


	\bibitem{Bao:2018zab} N.~Bao,
``Minimal Purifications, Wormhole Geometries, and the Complexity=Action Proposal,''
[arXiv:1811.03113 [hep-th]].





	\bibitem{Banados:1992wn} M.~Banados, C.~Teitelboim and J.~Zanelli,
``The Black hole in three-dimensional space-time,''
Phys. Rev. Lett. \textbf{69}, 1849-1851 (1992)
[arXiv:hep-th/9204099 [hep-th]].



	\bibitem{Gan:2016vuw} W.~C.~Gan, F.~W.~Shu and M.~H.~Wu,
``Emergent geometry, thermal CFT and surface/state correspondence,''
Phys. Lett. B \textbf{772}, 464-470 (2017)
[arXiv:1606.07628 [hep-th]].





\end{thebibliography}
\end{document}